\documentclass[12pt]{article}
\newcommand{\sech}{\mathrm{sech}}
\newcommand{\csch}{\mathrm{csch}}
\newcommand{\sn}{\mathrm{sn}}
\newcommand{\cn}{\mathrm{cn}}
\newcommand{\dn}{\mathrm{dn}}
\newcommand{\ds}{\mathrm{ds}}
\newcommand{\ns}{\mathrm{ns}}
\newcommand{\cs}{\mathrm{cs}}
\newcommand{\nd}{\mathrm{nd}}
\newcommand{\sd}{\mathrm{sd}}
\newcommand{\nc}{\mathrm{nc}}

\newtheorem{lemma}{Lemma}

\newtheorem{definition}{Definition}
\newtheorem{cas}{Case}
\begin{document}
\title{On classification of the solutions for general elliptic equation}
\author{Sirendaoreji\\{\small Mathematical Science College,Inner Mongolia
Normal University,}\\{\small Huhhot 010022,PR China}\\
{\small Email: siren@imnu.edu.cn}}
\date{October,11,2018}
\maketitle
{\bf Abstract:}\quad{The B\"acklund transformations and the superposition formulas for two sub--equations of the general elliptic equation are constructed from
the Riccati equation by using an indirect mapping method.
The thirty-six previously known solutions of the general elliptic
equation are proved to be equivalent to the another ten solutions.
The classification of solutions for the general elliptic equation is
obtained based on the equivalence relations.
The general elliptic equation expansion method with new classified
solutions are used to obtain abundant new exact traveling wave solutions
of a modified Camassa-Holm equation.}
\section{Introduction}
\label{sec:1}
Seeking exact solutions of nonlinear evolution equations (NLEEs) is a
practical and attractive problem in mathematical physics and nonlinear science.
People have made continuous effort on the problem and various direct methods
have been proposed to construct exact solutions for NLEEs. Among them the
Jacobi elliptic function expansion method~\cite{RefJ-1},the F--expansion method
~\cite{RefJ-2},the auxiliary equation method~\cite{RefJ-3} and the sub-equation method~\cite{RefJ-4,RefJ-5,RefJ-6} are related to the general elliptic
equation of the form~\cite{RefB-1,RefB-2,RefB-3,RefB-4,RefJ-7,RefJ-8}
\begin{equation}
\label{eq1:1}
{F^\prime}^2(\xi)=c_0+c_1F(\xi)+c_2F^2(\xi)+c_3F^3(\xi)+c_4F^4(\xi),
\end{equation}
where $c_i~(i=0,1,\cdots,4)$ are constants.Except for the sub--equation
method ,the Eq.(\ref{eq1:1}) was also chosen as the auxiliary equation in
other direct methods.It is the fact that the Eq.(\ref{eq1:1}) with
$c_0=1,c_1=0,c_2=-(m^2+1),c_3=0,c_4=m^2$
and $c_0=1-m^2,c_1=0,c_2=2m^2-1,c_3=0,c_4=-m^2$
were employed in the Jacobi elliptic function expansion method to seek the
snoidal wave solutions and the cnoidal wave solutions of NLEEs,respectively.
The F--expansion method for seeking twelve types of Jacobi
elliptic function solutions of NLEEs was proposed by using the
following sub-equation
\begin{equation}
\label{eq1:2}
{F^\prime}^2(\xi)=c_0+c_2F^2(\xi)+c_4F^4(\xi).
\end{equation}
Another sub--equation
~\cite{RefJ-9,RefJ-10,RefJ-11,RefJ-12,RefJ-13,RefJ-14,RefJ-15,RefJ-16,RefJ-17}
\begin{equation}
\label{eq1:3}
{F^\prime}^2(\xi)=c_2F^2(\xi)+c_3F^3(\xi)+c_4F^4(\xi),
\end{equation}
or
\begin{equation}
\label{eq1:4}
{G^\prime}^2(\xi)=h_2G^2(\xi)+h_4G^4(\xi)+h_6G^6(\xi),
\end{equation}
were used in the auxiliary equation method to find different
types of exact traveling wave solutions of NLEEs simultaneously.
Because of Eq.(\ref{eq1:3}) and Eq.(\ref{eq1:4}) can be transformed
into each other by the transformation
\begin{equation}
\label{eq1:5}
G(\xi)=F^{\frac 12}(\xi),h_2={\frac {c_2}4},h_4={\frac {c_3}4},h_6={\frac {c_4}4},
\end{equation}
so we only need to study Eq.(\ref{eq1:3}).In addition,the following
sub--equations~\cite{RefJ-18,RefJ-19,RefJ-20}
\begin{equation}
\label{eq1:6}
{F^\prime}^2(\xi)=c_0+c_1F(\xi)+c_2F^2(\xi)+c_3F^3(\xi),
\end{equation}
\begin{equation}
\label{eq1:7}
{F^\prime}^2(\xi)=c_0+c_1F(\xi)+c_2F^2(\xi),
\end{equation}
were also widely applied to seek exact traveling wave solutions of NLEEs.
The Eq.(\ref{eq1:1}) generalized to the case of involving six degree nonlinear term was
also studied by many authors~\cite{RefJ-21,RefJ-22,RefJ-23,RefJ-24,RefJ-25,RefJ-26,RefJ-27},
but according to the relation (\ref{eq1:5}) their results can be converted into
the case of Eq.(\ref{eq1:3}).
All these facts show that the Eq.(\ref{eq1:1}) has a strong application
background and its solutions are worth to further study. At present,
in the area of studying exact solutions of Eq.(\ref{eq1:1}),there exist
some problems such as
\begin{itemize}
\item[(1)] Many different solutions have been constructed and some of
them are found to be equivalent,but these equivalence relations have not
been proved.
\item[(2)] The classification of its solutions is not given.
\item[(3)] Its theoretical system such as the B\"acklund transformation (BT)
and the superposition formula (SF) has not been established.
\end{itemize}
For this reason,the paper aims to give a systematic study of
Eq.(\ref{eq1:1}) and try to solve the problems in a comprehensive way.
This effort lead us to organize the paper as follows.
In Sec.\ref{sec:2},we shall construct the BT and the SF for the
sub-equations of Eq.(\ref{eq1:1}).This lead us to construct the BT
and the SF of Eq.(\ref{eq1:3}) and Eq.(\ref{eq1:7}) from the
Riccati equation in terms of an indirect mapping method.
In Sec.\ref{sec:3},we shall give the proof
of the equivalence relations of the known solutions for Eq.(\ref{eq1:1}).
As a result,the thirty six previously known solutions are proved to be
equivalent to the another ten solutions for Eq.(\ref{eq1:1}).
In Sec.\ref{sec:4},we shall give the classification of the solutions
for Eq.(\ref{eq1:1}).It is shown that the solutions of Eq.(\ref{eq1:1})
are classified into five classes which consist of thirty eight independent solutions.
In Sec.\ref{sec:5},we shall construct abundant new exact traveling wave
solutions for the modified Camassa--Holm equation by means of the general
elliptic equation expansion method.Sec.\ref{sec:6} is devoted to concluding
remarks.
\section{B\"acklund transformations and superposition formulas}\label{sec:2}
Although the BT and the SF of Eq.(\ref{eq1:1}) is difficult to establish in a
simple way,but it usually can be constructed for its sub-equations through
use of some tedious integration procedure.For avoid the complexity,below
we shall construct the BT and the SF for Eq.(\ref{eq1:3}) and Eq.(\ref{eq1:7})
by using an indirect mapping method.
\subsection{The B\"acklund transformation and superposition formula of Eq.(\ref{eq1:3})}
Introducing the transformation
\begin{equation}
\label{eq2:1}
F(\xi)={\frac {4f^2(\xi)-c_2}{c_3-4\sqrt{c_4}f(\xi)}},
\end{equation}
which,substituted into the Eq.(\ref{eq1:3}),implies that
\begin{eqnarray*}
&\left({\frac {dF}{d\xi}}\right)^2-c_2F^2-c_3F^3-c_4F^4
  ={\frac {16\left({\frac {df}{d\xi}}-f^2+{\frac {c_2}{4}}\right)\left({\frac {df}{d\xi}}+f^2-{\frac {c_2}{4}}\right)H(f)}{\left(4\sqrt{c_4}f-c_3\right)^4}},\\
&H(f)=16c_4f^4-16c_3\sqrt{c_4}f^3+\left(4c_3^2+8c_2c_4\right)f^2-4c_2c_3\sqrt{c_4}f+c_2^2c_4.
\end{eqnarray*}
From this identity we deduce that $F(\xi)$ is a solution of Eq.(\ref{eq1:3}) if and only if
$f(\xi)$ is a solutions of the Riccati equation
\begin{equation}
\label{eq2:2}
f^\prime(\xi)=f^2(\xi)-{\frac 14}{c_2}.
\end{equation}
This shows that the BT and the SF of Eq.(\ref{eq1:3}) can be
induced from the BT and the SF of the Riccati equation (\ref{eq2:2}).
Using the BT and SF for Riccati equation~\cite{RefJ-28},
we obtain the BT and the SF of Eq.(\ref{eq2:2}) as follows
\begin{equation}
\label{eq2:3}
f_n={\frac {f_{n-1}+{\frac {c_2}{4}}\alpha}{1+\alpha{f_{n-1}}}}
={\frac {4f_{n-1}+\alpha{c_2}}{4+4\alpha{f_{n-1}}}},
\end{equation}
and
\begin{equation}
\label{eq2:4}
f_3={\frac {\alpha_1f_2-\alpha_2f_1}{\alpha_1f_1-\alpha_2f_2}}f_0,
\end{equation}
where $f_n,f_{n-1},f_i~(i=0,1,2,3)$ are solutions of
Eq.(\ref{eq2:2}) and $\alpha,\alpha_i~(i=1,2)$ are arbitrary constants.
Taking the mapping (\ref{eq2:1}) into consideration,we are confident
that two solutions $F_n$ and $F_{n-1}$ of Eq.(\ref{eq1:3}) can be written as
\begin{eqnarray}
\label{eq2:5}
&F_n={\frac {4f_n^2-c_2}{c_3-4\sqrt{c_4}f_n}},\\
\label{eq2:6}
&F_{n-1}={\frac {4f_{n-1}^2-c_2}{c_3-4\sqrt{c_4}f_{n-1}}}.
\end{eqnarray}
It solves from (\ref{eq2:6}) that
\begin{equation}
\label{eq2:7}
f_{n-1}=-{\frac 12}\sqrt{c_4}F_{n-1}+{\frac {\varepsilon}2}T_{n-1},
T_{n-1}=\sqrt{c_2+c_3F_{n-1}+c_4F_{n-1}^2},\varepsilon=\pm{1}.
\end{equation}
Therefore,substituting (\ref{eq2:3}) with (\ref{eq2:7}) into
(\ref{eq2:5}) leads the BT of Eq.(\ref{eq1:3}) of the form
\begin{equation}
\label{eq2:8}
F_n={\frac {H_{n-1}^2-c_2}{c_3-2\sqrt{c_4}H_{n-1}}},H_{n-1}={\frac {2(\varepsilon{T_{n-1}}-\sqrt{c_4}F_{n-1})+\alpha{c_2}}
    {2+\alpha(\varepsilon{T_{n-1}}-\sqrt{c_4}F_{n-1})}},
\end{equation}
where $T_{n-1}$ is determined by (\ref{eq2:7}).It is now easy to see
from (\ref{eq2:4}) and (\ref{eq2:5}) that the SF of Eq.(\ref{eq1:3})
is given by
\begin{equation}
\label{eq2:9}
\left\{\begin{array}{l}
F_{3a}={\frac {4f_3^2-c_2}{c_3-4\sqrt{c_4}f_3}},
   f_3={\frac {\alpha_1f_2-\alpha_2f_1}{\alpha_1f_1-\alpha_2f_2}}f_0,\\
f_n={\frac 12}\left(\varepsilon{T_n}-\sqrt{c_4}F_n\right),
   T_n=\sqrt{c_2+c_3F_n+c_4F_n^2},n=0,1,2.
\end{array}\right.
\end{equation}
\subsection{The B\"acklund transformation and superposition formula of Eq.(\ref{eq1:7})}
By virtue of the indirect mapping method as used above,we define a mapping
\begin{equation}
\label{eq2:10}
F(\xi)={\frac {c_1+4\sqrt{c_0}g(\xi)}{4g^2(\xi)-c_2}},
\end{equation}
and substitute it into Eq.(\ref{eq1:7}) leads
\begin{eqnarray*}
&\left({\frac {dF}{d\xi}}\right)^2-c_0-c_1F-c_2F^2
   ={\frac {16\left({\frac {dg}{d\xi}}-g^2+{\frac {c_2}{4}}\right)\left({\frac {dg}{d\xi}}+g^2-{\frac {c_2}{4}}\right)K(f)}{\left(4g^2-c_2\right)^4}},\\
&K(f)=16c_0g^4+16c_1\sqrt{c_0}g^3+\left(4c_1^2+8c_0c_2\right)g^2+4c_1c_2\sqrt{c_0}g+c_0c_2^2.
\end{eqnarray*}
This indicates that $F(\xi)$ is a solution of Eq.(\ref{eq1:7}) if and only if $g(\xi)$
satisfies the Riccati equation
\begin{equation}
\label{eq2:11}
g^\prime(\xi)=g^2(\xi)-{\frac 14}{c_2}.
\end{equation}
Following the same procedure as used to derive the BT and the SF for Eq.(\ref{eq2:2}),
we can now obtain the BT and the SF of Eq.(\ref{eq2:11})
\begin{equation}
\label{eq2:12}
g_n={\frac {4g_{n-1}+\beta{c_2}}{4+4\beta{g_{n-1}}}},
\end{equation}
and
\begin{equation}
\label{eq2:13}
g_3={\frac {\beta_1{g_2}-\beta_2{g_1}}{\beta_1{g_1}-\beta_2{g_2}}}g_0,
\end{equation}
respectively.It can be induced from the mapping (\ref{eq2:10}) that
\begin{eqnarray}
\label{eq2:14}
&F_n={\frac {c_1+4\sqrt{c_0}g_n}{4g_n^2-c_2}},\\
\label{eq2:15}
&F_{n-1}={\frac {c_1+4\sqrt{c_0}g_{n-1}}{4g_{n-1}^2-c_2}}.
\end{eqnarray}
At the same time,it solves from (\ref{eq2:15}) that
\begin{equation}
\label{eq2:16}
g_{n-1}={\frac 1{2F_{n-1}}}\left(\sqrt{c_0}+\varepsilon{R_{n-1}}\right),
R_{n-1}=\sqrt{c_0+c_1F_{n-1}+c_2F_{n-1}^2},\varepsilon=\pm{1}.
\end{equation}
The BT of Eq.(\ref{eq1:7}) cab be obtained by taking (\ref{eq2:12})
and (\ref{eq2:16}) into Eq.(\ref{eq2:14}),which is now determined to be
\begin{equation}
\label{eq2:17}
F_n={\frac {c_1+2\sqrt{c_0}K_{n-1}}{K_{n-1}^2-c_2}},
K_{n-1}={\frac {2\varepsilon{R_{n-1}}+\beta{c_2}F_{n-1}+2\sqrt{c_0}}
        {\beta\varepsilon{R_{n-1}}+2F_{n-1}+\beta\sqrt{c_0}}}.
\end{equation}
Finally,the superposition formula of Eq.(\ref{eq1:7}) is found to be
\begin{equation}
\label{eq2:18}
\left\{\begin{array}{l}
F_{3a}={\frac {c_1+4\sqrt{c_0}g_3}{4g_3^2-c_2}},
9_3={\frac {\beta_1{g_2}-\beta_2{g_1}}{\beta_1{g_1}-\beta_2{g_2}}}g_0,\\
g_n={\frac {\sqrt{c_0}+\varepsilon{R_n}}{2F_n}},R_n=\sqrt{c_0+c_1F_n+c_2F_n^2},n=0,1,2.
\end{array}\right.
\end{equation}
\section{Equivalence relations for solutions}\label{sec:3}
\begin{definition}
Two waves are called equivalence in the sense of wave translation if they have same waveform
but their phase difference is equal to a constant.Here this equivalence relation
will be expressed by the notation ``$\cong$".
\end{definition}
\begin{lemma}
\label{th3:1}
For any real numbers $\alpha$ and $\beta$,if $\alpha>{0},\beta>{0}$,then we hold that
\begin{eqnarray*}
\left\{\begin{array}{l}
\alpha{e^\xi}+\beta{e^{-\xi}}=2\sqrt{\alpha\beta}\cosh(\xi+{\frac 12}\ln{\frac \alpha\beta}),\\
\alpha{e^\xi}-\beta{e^{-\xi}}=2\sqrt{\alpha\beta}\sinh(\xi+{\frac 12}\ln{\frac \alpha\beta}).
\end{array}\right.
\end{eqnarray*}
\end{lemma}
\begin{lemma}
\label{th3:2}
For any real numbers $A$ and $B$,if $A^2>B^2$,then we have
\begin{eqnarray*}
\left\{\begin{array}{l}
A\sinh\eta\pm{B}\cosh\eta=\sqrt{A^2-B^2}\sinh\left(\eta\pm{\frac 12}\ln{\frac {A+B}{A-B}}\right),A>B,\\
A\sinh\eta\pm{B}\cosh\eta=-\sqrt{A^2-B^2}\sinh\left(\eta\pm{\frac 12}\ln{\frac {A+B}{A-B}}\right),A<B.
\end{array}\right.
\end{eqnarray*}
\end{lemma}
\begin{lemma}
\label{th3:3}
For any real numbers $A$ and $B$,if $A^2<B^2$,then we have
\begin{eqnarray*}
\left\{\begin{array}{l}
A\sinh\eta\pm{B}\cosh\eta=\mp\sqrt{B^2-A^2}\cosh\left(\eta\pm{\frac 12}\ln{\frac {B+A}{B-A}}\right),A>B,\\
A\sinh\eta\pm{B}\cosh\eta=\pm\sqrt{B^2-A^2}\cosh\left(\eta\pm{\frac 12}\ln{\frac {B+A}{B-A}}\right),A<B.
\end{array}\right.
\end{eqnarray*}
\end{lemma}
\begin{lemma}
\label{th3:4}
For any real numbers $A$ and $B$,we have
\begin{eqnarray*}
\left\{\begin{array}{l}
A\sin\eta\pm{B}\cos\eta=\sqrt{A^2+B^2}\sin\left(\eta\pm\theta\right),\\
A\cos\eta\pm{B}\sin\eta=\sqrt{A^2+B^2}\cos\left(\eta\mp\theta\right),
\end{array}\right.
\end{eqnarray*}
where $\theta=\arctan(B/A)$.
\end{lemma}
The above four lemmas are very important in classifying the traveling wave
solutions of NLEEs.The proof of the Lemma \ref{th3:1} and the Lemma \ref{th3:4}
are quite evidently.The Lemma \ref{th3:2} and the Lemma \ref{th3:3}
can be proved by using the Lemma \ref{th3:1}.So the proofs are omitted here.
\subsection{The equivalence relations for solutions of Eq.(\ref{eq1:3})}
We first prove the solutions of Eq.(\ref{eq1:3}) given by Table \ref{tab3:1}
are equivalent to the solutions given by Table \ref{tab3:2}.The
fourteen solutions were used in our previous studies~\cite{RefJ-10,RefJ-14}.
\begin{table}[htp]
\caption{Solutions of Eq.(\ref{eq1:3}) with $\Delta=c_3^2-4c_2c_4,\varepsilon=\pm{1}$.}
\label{tab3:1}
\begin{tabular}{llllll}
\hline\hline\noalign{\smallskip}
No& $z(\xi)$&Condition&{\quad}No& $z(\xi)$&Condition\\
\hline\noalign{\smallskip}
1&${\frac {-c_2c_3\sech^2\left({\frac {\sqrt{c_2}}2}\xi\right)}{c_3^2-c_2c_4\left(1+\varepsilon\tanh\left({\frac {\sqrt{c_2}}2}\xi\right)\right)^2}}$&$c_2>0$
&{\quad}5&${\frac {-c_2\sec^2\left({\frac {\sqrt{-c_2}}2}\xi\right)}{c_3+2\varepsilon\sqrt{-c_2c_4}\tan\left({\frac {\sqrt{-c_2}}2}\xi\right)}}$&$c_2<0,c_4>0$\\
2&${\frac {c_2c_3\csch^2\left({\frac {\sqrt{c_2}}2}\xi\right)}{c_3^2-c_2c_4\left(1+\varepsilon\coth\left({\frac {\sqrt{c_2}}2}\xi\right)\right)^2}}$&$c_2>0$
&{\quad}6&${\frac {-c_2\csc^2\left({\frac {\sqrt{-c_2}}2}\xi\right)}{c_3+2\varepsilon\sqrt{-c_2c_4}\cot\left({\frac {\sqrt{-c_2}}2}\xi\right)}}$&$c_2<0,c_4>0$\\
3&${\frac {-c_2\sech^2\left({\frac {\sqrt{c_2}}2}\xi\right)}{c_3+2\varepsilon\sqrt{c_2c_4}\tanh\left({\frac {\sqrt{c_2}}2}\xi\right)}}$&$c_2>0,c_4>0$
&{\quad}7&${\frac {4c_2{e^{\varepsilon\sqrt{c_2}\xi}}}{(e^{\varepsilon\sqrt{c_2}\xi}-c_3)^2-4c_2c_4}}$&$c_2>0$\\
4&${\frac {c_2\csch^2\left({\frac {\sqrt{c_2}}2}\xi\right)}{c_3+2\varepsilon\sqrt{c_2c_4}\coth\left({\frac {\sqrt{c_2}}2}\xi\right)}}$&$c_2>0,c_4>0$
&{\quad}8&${\frac {4\varepsilon{e^{\varepsilon\sqrt{c_2}\xi}}}{1-4c_2c_4e^{2\varepsilon\sqrt{c_2}\xi}}}$&$c_2>0,c_3=0$\\
\noalign{\smallskip}\hline
\end{tabular}
\end{table}
\begin{table}[htp]
\caption{Solutions of Eq.(\ref{eq1:3}) with $\Delta=c_3^2-4c_2c_4,\varepsilon=\pm{1}$.}
\label{tab3:2}
\begin{tabular}{llllll}
\hline\hline\noalign{\smallskip}
No& $F(\xi)$&Condition&{\quad}No & $F(\xi)$&Condition\\
\hline\noalign{\smallskip}
1&${\frac {2c_2}{\varepsilon\sqrt{\Delta}\cosh\left(\sqrt{c_2}\xi\right)-c_3}}$&$\Delta>0,c_2>0$
&{\quad}3b&${\frac {2c_2}{\varepsilon\sqrt{\Delta}\sin\left(\sqrt{-c_2}\xi\right)-c_3}}$&$\Delta>0,c_2<0$\\
2&${\frac {2c_2}{\varepsilon\sqrt{-\Delta}\sinh\left(\sqrt{c_2}\xi\right)-c_3}}$&$\Delta<0,c_2>0$
&{\quad}4&$-{\frac {c_2}{c_3}}\left[1+\varepsilon\tanh\left({\frac {\sqrt{c_2}}2}\xi\right)\right]$&$\Delta=0,c_2>0$\\
3a&${\frac {2c_2}{\varepsilon\sqrt{\Delta}\cos\left(\sqrt{-c_2}\xi\right)-c_3}}$&$\Delta>0,c_2<0$
&{\quad}5&$-{\frac {c_2}{c_3}}\left[1+\varepsilon\coth\left({\frac {\sqrt{c_2}}2}\xi\right)\right]$&$\Delta=0,c_2>0$\\
\noalign{\smallskip}\hline
\end{tabular}
\end{table}
\par
Let $A=2c_2c_4,B=c_3^2-2c_2c_4$,then we find $A^2-B^2=-c_3^2\Delta$.Hence the
condition $\Delta>0$ implies that $A^2<B^2$ and $A<B$.But for condition $\Delta<0$
we obtain that $A^2>B^2$ and $A>B$.Therefore,noting the definition of hyperbolic
functions,using the formulas
\begin{equation}
\label{eq3:1}
\sinh{x}=2\sinh{\frac x2}\cosh{\frac x2},\cosh^2{\frac x2}={\frac {\cosh{x}+1}2},
\sinh^2{\frac x2}={\frac {\cosh{x}-1}2},
\end{equation}
the Lemma \ref{th3:3} and the Lemma \ref{th3:2} we can derive that
\begin{eqnarray*}
\begin{array}{l}
z_1(\xi)={\frac {-2c_2c_3}{c_3^2\mp\left[2c_2c_4\sinh\left(\sqrt{c_2}\xi\right)\right.
  \left.\mp\left(c_3^2-2c_2c_4\right)\cosh\left(\sqrt{c_2}\xi\right)\right]}},\\
\quad=\left\{\begin{array}{l}
   {\frac {2c_2}{\varepsilon\sqrt{\Delta}\cosh\left(\sqrt{c_2}\xi\mp{\frac 12}\ln({\frac {c_3^2}{\Delta}})\right)-c_3}}\cong{F_1(\xi)},\Delta>0,c_2>0,\\
   {\frac {2c_2}{\varepsilon\sqrt{-\Delta}\sinh\left(\sqrt{c_2}\xi\mp{\frac 12}\ln({\frac {c_3^2}{-\Delta}})\right)-c_3}}\cong{F_2(\xi)},\Delta<0,c_2>0.
\end{array}\right.
\end{array}
\end{eqnarray*}
Following the same procedure we can derive that $z_2(\xi)\cong{F_1(\xi)}$,for $\Delta>0,c_2>0$ and $z_2(\xi)\cong{F_2(\xi)}$,for $\Delta<0,c_2>0$ and
the details of the proof is omitted here.\newline
Set $A=2\sqrt{c_2c_4},B=c_3$,then $A^2-B^2=-\Delta$.Thus for $\Delta>0$
we hold $A^2<B^2$.But for $\Delta<0$,we have $A^2>B^2$.Hence for $c_2>0,c_4>0$,
the formulas (\ref{eq3:1}),the Lemma \ref{th3:2} and the Lemma \ref{th3:3}
are used to obtain that
\begin{eqnarray*}
\begin{array}{l}
z_3(\xi)={\frac {-2c_2}{c_3\pm\left[2\sqrt{c_2c_4}\sinh\left(\sqrt{c_2}\xi\right)\pm{c_3}\cosh\left(\sqrt{c_2}\xi\right)\right]}},\\
\quad=\left\{\begin{array}{l}
   {\frac {2c_2}{\varepsilon\sqrt{\Delta}\cosh\left(\sqrt{c_2}\xi\pm{\frac 12}\ln({\frac {c_3+2\sqrt{c_2c_4}}{c_3-2\sqrt{c_2c_4}}})\right)-c_3}}\cong{F_1(\xi)},\Delta>0,\\
   {\frac {2c_2}{\varepsilon\sqrt{-\Delta}\sinh\left(\sqrt{c_2}\xi\pm{\frac 12}\ln({\frac {2\sqrt{c_2c_4}+c_3}{2\sqrt{c_2c_4}-c_3}})\right)-c_3}}\cong{F_2(\xi)},\Delta<0.
   \end{array}\right.
\end{array}
\end{eqnarray*}
Applying the same procedure as used to prove $z_3(\xi)$,we obtain the results
$z_4(\xi)\cong{F_1(\xi)},\Delta>0$ and $z_4(\xi)\cong{F_2(\xi)},\Delta<0$.\par
If $\Delta=0$,then $4c_2c_4=c_3^2$.In this case,for $c_2>0,c_4>0$ we find
\begin{eqnarray*}
\begin{array}{l}
z_3(\xi)={\frac {-c_2\left(1-\tanh^2\left({\frac {\sqrt{c_2}}2}\xi\right)\right)}{c_3+\varepsilon{c_3}\tanh\left({\frac {\sqrt{c_2}}2}\xi\right)}}=-{\frac {c_2}{c_3}}\left[1+\varepsilon\tanh\left({\frac {\sqrt{c_2}}2}\xi\right)\right]=F_4(\xi),\\
z_4(\xi)={\frac {-c_2\left(1-\coth^2\left({\frac {\sqrt{c_2}}2}\xi\right)\right)}{c_3+\varepsilon{c_3}\coth\left({\frac {\sqrt{c_2}}2}\xi\right)}}=-{\frac {c_2}{c_3}}\left[1+\varepsilon\coth\left({\frac {\sqrt{c_2}}2}\xi\right)\right]=F_5(\xi).
\end{array}
\end{eqnarray*}
When $\Delta>0,c_2<0,c_4>0$,using the expressions $\sec{x}={\frac 1{\cos{x}}},\csc{x}={\frac 1{\sin{x}}}$,the identities
\begin{equation}
\label{eq3:2}
\sin{x}=2\sin{\frac x2}\cos{\frac x2},
\sin^2{\frac x2}={\frac {1-\cos{x}}2},
\cos^2{\frac x2}={\frac {1+\cos{x}}2},
\end{equation}
and the Lemma \ref{th3:4},the solutions $z_5(\xi)$ can be simplified as
\begin{eqnarray*}
\begin{array}{l}
z_5(\xi)={\frac {-2c_2}{2\varepsilon\sqrt{-c_2c_4}\sin\left(\sqrt{-c_2}\xi\right)+{c_3}\cos\left(\sqrt{-c_2}\xi\right)+c_3}},\\
\quad=\left\{\begin{array}{l}
{\frac {2c_2}{-\sqrt{\Delta}\cos\left(\sqrt{-c_2}\xi\mp\theta_1\right)-c_3}}
   {\cong}F_{3a}(\xi)|_{\varepsilon=-1},\\
{\frac {2c_2}{\varepsilon\sqrt{\Delta}\sin\left(\sqrt{-c_2}\xi\pm\theta_2\right)-c_3}}
   {\cong}F_{3b}(\xi),
\end{array}\right.
\end{array}
\end{eqnarray*}
where $\theta_1=\arctan\left({\frac {2\sqrt{-c_2c_4}}{c_3}}\right),
\theta_2=\arctan\left({\frac {c_3}{2\sqrt{-c_2c_4}}}\right)$.
Similarly,when $\Delta>0,c_2<0,c_4>0$,we can prove that $z_6(\xi){\cong}F_{3a}(\xi)|_{\varepsilon=-1}$ and $z_6(\xi){\cong}F_{3b}(\xi)$.
\par
Converting $z_7(\xi)$ into the exponential form and then using
the Lemma \ref{th3:1},we obtain
\begin{eqnarray*}
\begin{array}{l}
z_7(\xi)={\frac {4c_2}{e^{\varepsilon\sqrt{c_2}\xi}+(c_3^2-4c_2c_4)
    e^{-\varepsilon\sqrt{c_2}\xi}-2c_3}},\\
    \quad=\left\{\begin{array}{l}
      {\frac {2c_2}{\sqrt{\Delta}\cosh\left(\varepsilon\sqrt{c_2}\xi
       +{\frac 12}\ln({\frac 1\Delta})\right)-c_3}},\\
       {\frac {2c_2}{\sqrt{-\Delta}\sinh\left(\varepsilon\sqrt{c_2}\xi+{\frac 12}\ln({\frac 1{-\Delta}})\right)-c_3}},
\end{array}\right.\\
\quad=\left\{\begin{array}{l}
{\frac {2c_2}{\sqrt{\Delta}\cosh\left(\sqrt{c_2}\xi\pm{\frac 12}\ln({\frac 1\Delta})\right)-c_3}}\cong{F_1(\xi)|_{\varepsilon=1},\Delta>0,c_2>0},\\
{\frac {2c_2}{\varepsilon\sqrt{-\Delta}\sinh\left(\sqrt{c_2}\xi\pm{\frac 12}\ln({\frac 1{-\Delta}})\right)-c_3}}\cong{F_2(\xi)},\Delta<0,c_2>0.
\end{array}\right.
\end{array}
\end{eqnarray*}
When $\Delta=0,c_2>0$,it is obtained from the identities ${\frac 1{1+e^x}}={\frac 12}\left(1-\tanh{\frac x2}\right)$
and ${\frac 1{1-e^x}}={\frac 12}\left(1-\coth{\frac x2}\right)$ that
\begin{eqnarray*}
\begin{array}{l}
z_7(\xi)={\frac {4c_2}{e^{\varepsilon\sqrt{c_2}\xi}-2c_3}}=\left\{\begin{array}{l}
  -{\frac {2c_2}{c_3}}{\frac 1{1-e^{\varepsilon\sqrt{c_2}\xi+\ln({\frac 1{2c_3}})}}},\\
   -{\frac {2c_2}{c_3}}{\frac 1{1+e^{\varepsilon\sqrt{c_2}\xi+\ln({\frac {-1}{2c_3}})}}},\\
\end{array}\right.\\
\quad=\left\{\begin{array}{l}
    -{\frac {c_2}{c_3}}\left[1-\coth\left({\frac {\varepsilon\sqrt{c_2}}{2}\xi+\ln({\frac 1{2c_3}})}\right)\right],\\
     -{\frac {c_2}{c_3}}\left[1-\tanh\left({\frac {\varepsilon\sqrt{c_2}}{2}\xi+\ln({\frac {-1}{2c_3}})}\right)\right],\\
\end{array}\right.\\
\quad=\left\{\begin{array}{l}
-{\frac {c_2}{c_3}}\left[1+\varepsilon\coth\left({\frac {\sqrt{c_2}}{2}\xi\pm\ln({\frac 1{2c_3}})}\right)\right]\cong{F_5(\xi)},c_3>0,\\
-{\frac {c_2}{c_3}}\left[1+\varepsilon\tanh\left({\frac {\sqrt{c_2}}{2}\xi\pm\ln({\frac {-1}{2c_3}})}\right)\right]\cong{F_4(\xi)},c_3<0.
\end{array}\right.
\end{array}
\end{eqnarray*}
For $c_2>0,c_3=0$,it yields from the Lemma \ref{th3:1} that
\begin{eqnarray*}
\begin{array}{l}
z_8(\xi)={\frac {4\varepsilon{c_2}}{e^{-\varepsilon\sqrt{c_2}\xi}-4c_2c_4e^{\varepsilon\sqrt{c_2}\xi}}}
  =\left\{\begin{array}{l}
{\frac {4\varepsilon{c_2}}{2\sqrt{-4c_2c_4}\cosh\left(\varepsilon\sqrt{c_2}\xi+{\frac 12}\ln(-4c_2c_4)\right)}},\\
{\frac {4\varepsilon{c_2}}{-2\sqrt{4c_2c_4}\sinh\left(\varepsilon\sqrt{c_2}\xi+{\frac 12}\ln(4c_2c_4)\right)}},
\end{array}\right.\\
\quad=\left\{\begin{array}{l}
\varepsilon\sqrt{-\frac {c_2}{c_4}}\sech\left(\sqrt{c_2}\xi\pm{\frac 12}\ln(-4c_2c_4)\right)
\cong{F_1(\xi),c_4<0},\\
\varepsilon\sqrt{\frac {c_2}{c_4}}\csch\left(\sqrt{c_2}\xi\pm{\frac 12}\ln(4c_2c_4)\right)
\cong{F_2(\xi),c_4>0.}
\end{array}\right.
\end{array}
\end{eqnarray*}
The above obtained results on the equivalence relations for solutions
of Eq.(\ref{eq1:3}) are consistent with the Liu's results~\cite{RefJ-29,RefJ-30},
but our proofs are more simple and clear than that of Liu's proofs because
the proving processes have been simplified by the above four Lemmas.\par
We next consider the following four solutions~\cite{RefJ-31}
\begin{eqnarray*}
\begin{array}{l}
\phi_5(\xi)={\frac {-2c_2\sech\left(\sqrt{c_2}\xi\right)}{c_3\sech\left(\sqrt{c_2}\xi\right)
   +\sqrt{4c_2^2-c_3^2+4c_2c_4}\tanh\left(\sqrt{c_2}\xi\right)-2c_2}},c_2>0,\\
\phi_6(\xi)={\frac {-2c_2\csch\left(\sqrt{c_2}\xi\right)}{c_3\csch\left(\sqrt{c_2}\xi\right)
   -\sqrt{4c_2^2+c_3^2-4c_2c_4}\coth\left(\sqrt{c_2}\xi\right)+2c_2}},c_2>0,\\
\phi_7(\xi)={\frac {-2c_2\sec\left(\sqrt{-c_2}\xi\right)}{c_3\sec\left(\sqrt{-c_2}\xi\right)
   +\sqrt{c_3^2-4c_2^2-4c_2c_4}\tan\left(\sqrt{-c_2}\xi\right)-2c_2}},c_2<0,\\
\phi_8(\xi)={\frac {-2c_2\csc\left(\sqrt{-c_2}\xi\right)}{c_3\csc\left(\sqrt{-c_2}\xi\right)
   -\sqrt{c_3^2-4c_2^2-4c_2c_4}\cot\left(\sqrt{-c_2}\xi\right)+2c_2}},c_2<0.
\end{array}
\end{eqnarray*}
Let $A=\sqrt{4c_2^2-c_3^2+4c_2c_4},B=2c_2$,then we get $A^2-B^2=-\Delta$.
However,for $c_2>0$,the condition $\Delta>0$ yields that $A^2<B^2$ and $A<B$.
And the condition $\Delta<0$ gives that $A^2>B^2$ and $A>B$.
Therefore,when $c_2>0$,it follows from the Lemma \ref{th3:3} and the Lemma \ref{th3:2} that
\begin{eqnarray*}
\begin{array}{l}
\phi_5(\xi)={\frac {-2c_2}{c_3-2c_2\cosh\left(\sqrt{c_2}\xi\right)
   +\sqrt{4c_2^2-c_3^2+4c_2c_4}\sinh\left(\sqrt{c_2}\xi\right)}},\\
\quad=\left\{\begin{array}{l}
  {\frac {2c_2}{\sqrt{\Delta}\cosh\left(\sqrt{c_2}\xi-{\frac 12}\ln(\theta_1)\right)-c_3}}
  \cong{F_1(\xi)}|_{\varepsilon=1},\Delta>0,\\
  {\frac {2c_2}{-\sqrt{-\Delta}\sinh\left(\sqrt{c_2}\xi-{\frac 12}\ln(\theta_2)\right)-c_3}}
  \cong{F_2(\xi)}|_{\varepsilon=-1},\Delta<0.
\end{array}\right.
\end{array}
\end{eqnarray*}
where
\begin{eqnarray*}
\theta_1={\frac {2c_2+\sqrt{4c_2^2-c_3^2+4c_2c_4}}{2c_2-\sqrt{4c_2^2-c_3^2+4c_2c_4}}},
\theta_2={\frac {\sqrt{4c_2^2-c_3^2+4c_2c_4}+2c_2}{\sqrt{4c_2^2-c_3^2+4c_2c_4}-2c_2}}.
\end{eqnarray*}
By introducing $A=2c_2,B=\sqrt{4c_2^2+c_3^2-4c_2c_4}$,we find that $A^2-B^2=-\Delta$.
When $c_2>0$,the condition $\Delta>0$ implies that $A^2<B^2$ and
$A<B$.On the other hand,from the condition $\Delta<0$ we can deduce
that $A^2>B^2$ and $A>B$.Hence for $c_2>0$,it is obtained from the
Lemma \ref{th3:3} and the Lemma \ref{th3:2} that
\begin{eqnarray*}
\begin{array}{l}
\phi_6(\xi)={\frac {-2c_2}{c_3+{2c_2}\sinh\left(\sqrt{c_2}\xi\right)
   -\sqrt{4c_2^2+c_3^2-4c_2c_4}\cosh\left(\sqrt{c_2}\xi\right)}},\\
\quad=\left\{\begin{array}{l}
    {\frac {2c_2}{\sqrt{\Delta}\cosh\left(\sqrt{c_2}\xi-{\frac 12}\ln(\theta_1)\right)-c_3}}
    \cong{F_1(\xi)}|_{\varepsilon=1},\Delta>0,\\
    {\frac {2c_2}{-\sqrt{-\Delta}\sinh\left(\sqrt{c_2}\xi-{\frac 12}\ln(\theta_2)\right)-c_3}}
  \cong{F_2(\xi)}|_{\varepsilon=-1},\Delta<0.
\end{array}\right.
\end{array}
\end{eqnarray*}
where
\begin{eqnarray*}
\theta_1={\frac {\sqrt{4c_2^2+c_3^2-4c_2c_4}+2c_2}{\sqrt{4c_2^2+c_3^2-4c_2c_4}-2c_2}},
\theta_2={\frac {2c_2+\sqrt{4c_2^2+c_3^2-4c_2c_4}}{2c_2-\sqrt{4c_2^2+c_3^2-4c_2c_4}}}.
\end{eqnarray*}
For $\Delta>0,c_2<0$ we obtain from the Lemma \ref{th3:4} that
\begin{eqnarray*}
\begin{array}{l}
\phi_7(\xi)={\frac {-2c_2}{c_3-2c_2\cos\left(\sqrt{-c_2}\xi\right)
    +\sqrt{c_3^2-4c_2^2-4c_2c_4}\sin\left(\sqrt{-c_2}\xi\right)}},\\
\quad=\left\{\begin{array}{l}
  {\frac {2c_2}{\sqrt{\Delta}\cos\left(\sqrt{-c_2}\xi+\theta_1\right)-c_3}}
  {\cong}F_{3a}(\xi)|_{\varepsilon=1},\\
  {\frac {2c_2}{-\sqrt{\Delta}\sin\left(\sqrt{-c_2}\xi-\theta_2\right)-c_3}}
  {\cong}F_{3b}(\xi)|_{\varepsilon=-1},
\end{array}\right.\\
\phi_8(\xi)={\frac {-2c_2}{c_3+2c_2\sin\left(\sqrt{-c_2}\xi\right)
    -\sqrt{c_3^2-4c_2^2-4c_2c_4}\cos\left(\sqrt{-c_2}\xi\right)}},\\
\quad=\left\{\begin{array}{l}
  {\frac {2c_2}{\sqrt{\Delta}\cos\left(\sqrt{-c_2}\xi+\theta_1\right)-c_3}}{\cong}F_{3a}(\xi)|_{\varepsilon=1},\\
  {\frac {2c_2}{-\sqrt{\Delta}\sin\left(\sqrt{-c_2}\xi-\theta_2\right)-c_3}}{\cong}F_{3b}(\xi)|_{\varepsilon=-1},
\end{array}\right.
\end{array}
\end{eqnarray*}
where
\begin{eqnarray*}
\theta_1=\arctan\left({\frac {\sqrt{c_3^2-4c_2^2-4c_2c_3}}{2c_2}}\right),
\theta_2=\arctan\left({\frac {2c_2}{\sqrt{c_3^2-4c_2^2-4c_2c_3}}}\right).
\end{eqnarray*}
Third we turn to consider the eight solutions listed in Table \ref{tab3:3}
which were obtained by Yang {\it et al}~\cite{RefJ-32}.\newpage
\begin{table}[htp]
\caption{Solutions of Eq.(\ref{eq1:3}) with $\Delta=c_3^2-4c_2c_4>0$.}
\label{tab3:3}
\begin{tabular}{llllll}
\hline\hline\noalign{\smallskip}
No & $\varphi(\xi)$&Condition&{\quad}No & $\varphi(\xi)$&Condition \\
\noalign{\smallskip}\hline\noalign{\smallskip}
1&${\frac {2c_2\sech^2\left({\frac {\sqrt{c_2}}{2}}\xi\right)}{2\sqrt{\Delta}-(\Delta+c_3)\sech^2\left({\frac {\sqrt{c_2}}{2}}\xi\right)}}$&$c_2>0$
&{\quad}5&${\frac {2c_2\sec^2\left({\frac {\sqrt{-c_2}}{2}}\xi\right)}{2\sqrt{\Delta}-(\Delta+c_3)\sec^2\left({\frac {\sqrt{-c_2}}{2}}\xi\right)}}$&$c_2<0$\\
2&${\frac {2c_2\sech^2\left({\frac {\sqrt{c_2}}{2}}\xi\right)}{-2\sqrt{\Delta}+(\Delta-c_3)\sech^2\left({\frac {\sqrt{c_2}}{2}}\xi\right)}}$&$c_2>0$
&{\quad}6&${\frac {2c_2\sec^2\left({\frac {\sqrt{-c_2}}{2}}\xi\right)}{-2\sqrt{\Delta}+(\Delta-c_3)\sec^2\left({\frac {\sqrt{-c_2}}{2}}\xi\right)}}$&$c_2<0$\\
3&${\frac {2c_2\csch^2\left({\frac {\sqrt{c_2}}{2}}\xi\right)}{2\sqrt{\Delta}+(\Delta-c_3)\csch^2\left({\frac {\sqrt{c_2}}{2}}\xi\right)}}$&$c_2>0$
&{\quad}7&${\frac {2c_2\csc^2\left({\frac {\sqrt{-c_2}}{2}}\xi\right)}{2\sqrt{\Delta}-(\Delta+c_3)\csc^2\left({\frac {\sqrt{-c_2}}{2}}\xi\right)}}$&$c_2<0$\\
4&${\frac {2c_2\csch^2\left({\frac {\sqrt{c_2}}{2}}\xi\right)}{-2\sqrt{\Delta}-(\Delta+c_3)\csch^2\left({\frac {\sqrt{c_2}}{2}}\xi\right)}}$&$c_2>0$
&{\quad}8&${\frac {2c_2\csc^2\left({\frac {\sqrt{-c_2}}{2}}\xi\right)}{-2\sqrt{\Delta}+(\Delta-c_3)\csc^2\left({\frac {\sqrt{-c_2}}{2}}\xi\right)}}$&$c_2<0$\\
\noalign{\smallskip}\hline
\end{tabular}
\end{table}
\par\noindent
Using the definition of the hyperbolic functions and trigonometric functions,
the formulas (\ref{eq3:1}) and (\ref{eq3:2}),we find that
\begin{eqnarray*}
\begin{array}{l}
\varphi_1(\xi)={\frac {2c_2}{2\sqrt{\Delta}\cosh^2\left({\frac {\sqrt{c_2}}{2}}\xi\right)-\sqrt{\Delta}-c_3}}
  ={\frac {2c_2}{\sqrt{\Delta}\cosh\left(\sqrt{c_2}\xi\right)-c_3}}
=F_1(\xi)|_{\varepsilon=1},\\
\varphi_3(\xi)={\frac {2c_2}{2\sqrt{\Delta}\sinh^2\left({\frac {\sqrt{c_2}}{2}}\xi\right)+\sqrt{\Delta}-c_3}}
   ={\frac {2c_2}{\sqrt{\Delta}\cosh\left(\sqrt{c_2}\xi\right)-c_3}}
=F_1(\xi)|_{\varepsilon=1},\\
\varphi_5(\xi)={\frac {2c_2}{2\sqrt{\Delta}\cos^2\left({\frac {\sqrt{-c_2}}{2}}\xi\right)-\sqrt{\Delta}-c_3}}
   ={\frac {2c_2}{\sqrt{\Delta}\cos\left(\sqrt{-c_2}\xi\right)-c_3}}
   =F_{3a}(\xi)|_{\varepsilon=1},\\
\varphi_8(\xi)={\frac {2c_2}{-2\sqrt{\Delta}\sin^2\left({\frac {\sqrt{-c_2}}{2}}\xi\right)+\sqrt{\Delta}-c_3}}
  ={\frac {2c_2}{\sqrt{\Delta}\cos\left(\sqrt{-c_2}\xi\right)-c_3}}
  =F_{3a}(\xi)|_{\varepsilon=1}.
\end{array}
\end{eqnarray*}
Proceeding as before, we can easily prove that $\varphi_2(\xi)={F_1(\xi)}|_{\varepsilon=-1},\varphi_4(\xi)={F_1(\xi)}|_{\varepsilon=-1},
\varphi_6(\xi)={F_{3a}(\xi)}|_{\varepsilon=-1},\varphi_7(\xi)={F_{3a}(\xi)}|_{\varepsilon=-1}$.
\par\noindent
Corresponding to the case $c_4=0$, we shall now prove the following
two solutions~\cite{RefJ-33} of Eq.(\ref{eq1:3})
\begin{eqnarray*}
\begin{array}{l}
f_1(\xi)=-{\frac {4c_2\left(\cosh\left(\sqrt{c_2}\xi\right)+\sinh\left(\sqrt{c_2}\xi\right)\right)}
{\left(b+\cosh\left(\sqrt{c_2}\xi\right)+\sinh\left(\sqrt{c_2}\xi\right)\right)^2}},c_2>0,\\
f_2(\xi)={\frac {4c_2\left(\cosh\left(\sqrt{c_2}\xi\right)+\sinh\left(\sqrt{c_2}\xi\right)\right)}
 {\left(-b+\cosh\left(\sqrt{c_2}\xi\right)+\sinh\left(\sqrt{c_2}\xi\right)\right)^2}},c_2>0,
\end{array}
\end{eqnarray*}
satisfy the equivalent relation $f_1(\xi)\cong{F_1(\xi)}|_{\varepsilon=-1}$ and
$f_2(\xi)\cong{F_1(\xi)}|_{\varepsilon=1}$.\par
For $c_4=0$,using the identity $\sinh{x}+\cosh{x}=e^x$ and the Lemma \ref{th3:1} yields
\begin{eqnarray*}
\begin{array}{l}
f_1(\xi)=-{\frac {4c_2}{\left(e^{{\frac {\sqrt{c_2}}2}\xi}+c_3e^{-{\frac {\sqrt{c_2}}2}\xi}\right)^2}}=\left\{\begin{array}{l}
  {\frac {-4c_2}{\left[2\sqrt{c_3}\cosh\left({\frac {\sqrt{c_2}}2}\xi-\ln(\sqrt{c_3})\right)\right]^2}},\\
  {\frac {-4c_2}{\left[2\sqrt{-c_3}\sinh\left({\frac {\sqrt{c_2}}2}\xi-\ln(\sqrt{-c_3})\right)\right]^2}},
    \end{array}\right.\\
\quad\quad\;=\left\{\begin{array}{l}
  -{\frac {c_2}{c_3}}\sech^2\left({\frac {\sqrt{c_2}}{2}}\xi-\ln(\sqrt{c_3})\right),c_3>0,\\
  {\frac {c_2}{c_3}}\csch^2\left({\frac {\sqrt{c_2}}{2}}\xi-\ln(\sqrt{-c_3})\right),c_3<0,
\end{array}\right.
\end{array}
\end{eqnarray*}
\begin{eqnarray*}
\begin{array}{l}
f_2(\xi)={\frac {4c_2}{\left(e^{{\frac {\sqrt{c_2}}2}\xi}-c_3e^{-{\frac {\sqrt{c_2}}2}\xi}\right)^2}},=\left\{\begin{array}{l}
   {\frac {4c_2}{\left[2\sqrt{c_3}\sinh\left({\frac {\sqrt{c_2}}2}\xi-\ln(\sqrt{c_3})\right)\right]^2}},\\
   {\frac {4c_2}{\left[2\sqrt{-c_3}\cosh\left({\frac {\sqrt{c_2}}2}\xi-\ln(\sqrt{-c_3})\right)\right]^2}},
\end{array}\right.\\
\quad\quad\;=\left\{\begin{array}{l}
   {\frac {c_2}{c_3}}\csch^2\left({\frac {\sqrt{c_2}}{2}}\xi-\ln(\sqrt{c_3})\right),c_3>0,\\
   -{\frac {c_2}{c_3}}\sech^2\left({\frac {\sqrt{c_2}}{2}}\xi-\ln(\sqrt{-c_3})\right),c_3<0.
\end{array}\right.
\end{array}
\end{eqnarray*}
These are just the results we want to prove.
\subsection{The equivalence relations for solutions of Eq.(\ref{eq1:7})}
Finally,we shall prove the twelve solutions~\cite{RefJ-34} of Eq.(\ref{eq1:7})
listed in Table \ref{tab3:4}
\begin{table*}[!htp]
\caption{Solutions of Eq.(\ref{eq1:7}).}
\label{tab3:4}
\begin{tabular}{lll}
\hline\hline\noalign{\smallskip}
No & $Z(\xi)$ &Condition\\
\noalign{\smallskip}\hline
1 &$ {\frac 1{c_2}}\left[c_1-2\sqrt{c_0c_2}\coth\left({\frac {\sqrt{c_2}}{2}}\xi\right)\right]\sinh^2\left({\frac {\sqrt{c_2}}{2}}\xi\right)$&$c_0>0,c_2>0$ \\
2 &$ -{\frac 1{c_2}}\left[c_1-2\sqrt{c_0c_2}\tanh\left({\frac {\sqrt{c_2}}{2}}\xi\right)\right]\cosh^2\left({\frac {\sqrt{c_2}}{2}}\xi\right)$&$c_0>0,c_2>0$\\
3&$-{\frac 1{c_2}}\left[c_1+2\sqrt{-c_0c_2}\tan\left({\frac {\sqrt{-c_2}}{2}}\xi\right)\right]\cos^2\left({\frac {\sqrt{-c_2}}{2}}\xi\right)$&$c_0>0,c_2<0$\\
4&$-{\frac 1{c_2}}\left[c_1-2\sqrt{-c_0c_2}\cot\left({\frac {\sqrt{-c_2}}{2}}\xi\right)\right]\sin^2\left({\frac {\sqrt{-c_2}}{2}}\xi\right)$&$c_0>0,c_2<0$\\
5&${\frac 1{4c_2}}\left[-2c_1-\left(1+c_1^2-4c_0c_2\right)\cosh\left(\sqrt{c_2}\xi\right)\right.$&\quad\\
\quad&\quad$+\left.\left(1-c_1^2+4c_0c_2\right)\sinh\left(\sqrt{c_2}\xi\right)\right]$&$c_2>0$\\
6&$-{\frac 1{2c_2}}+{\frac {c_1^2-4c_0c_2+4c_2^2\left(\cosh\left(\sqrt{c_2}\xi\right)
   +\sinh\left(\sqrt{c_2}\xi\right)\right)^2}{8c_2^2\left(\cosh\left(\sqrt{c_2}\xi\right)
   +\sinh\left(\sqrt{c_2}\xi\right)\right)}}$&$c_2>0$\\[0.25cm]
7&$-{\frac {c_1+2\sqrt{-c_0c_2}\left(\tan\left(\sqrt{-c_2}\xi\right)
   +\sec\left(\sqrt{-c_2}\xi\right)\right)}{c_2\left[1+\left(\tan\left(\sqrt{-c_2}\xi\right)
   +\sec\left(\sqrt{-c_2}\xi\right)\right)^2\right]}}$&$c_2<0,c_0>0$\\[0.25cm]
8&$-{\frac {c_1+2\sqrt{-c_0c_2}\left(\tan\left(\sqrt{-c_2}\xi\right)
   -\sec\left(\sqrt{-c_2}\xi\right)\right)}{c_2\left[1+\left(\tan\left(\sqrt{-c_2}\xi\right)
   -\sec\left(\sqrt{-c_2}\xi\right)\right)^2\right]}}$&$c_2<0,c_0>0$\\[0.25cm]
9&${\frac {-c_1+2\sqrt{-c_0c_2}\left(\cot\left(\sqrt{-c_2}\xi\right)
   +\csc\left(\sqrt{-c_2}\xi\right)\right)}{c_2\left[1+\left(\cot\left(\sqrt{-c_2}\xi\right)
   +\csc\left(\sqrt{-c_2}\xi\right)\right)^2\right]}}$&$c_2<0,c_0>0$\\[0.25cm]
10&${\frac {-c_1+2\sqrt{-c_0c_2}\left(\cot\left(\sqrt{-c_2}\xi\right)
   -\csc\left(\sqrt{-c_2}\xi\right)\right)}{c_2\left[1+\left(\cot\left(\sqrt{-c_2}\xi\right)
   -\csc\left(\sqrt{-c_2}\xi\right)\right)^2\right]}}$&$c_2<0,c_0>0$\\[0.25cm]
11&${\frac {c_1-2\sqrt{c_0c_2}\left(\tanh\left(\sqrt{c_2}\xi\right)
   +i\sech\left(\sqrt{c_2}\xi\right)\right)}{-c_2\left[1-\left(\tanh\left(\sqrt{c_2}\xi\right)
   +i\sech\left(\sqrt{c_2}\xi\right)\right)^2\right]}}$&$c_2>0,c_0>0$\\[0.25cm]
12&${\frac {c_1-2\sqrt{c_0c_2}\left(\tanh\left(\sqrt{c_2}\xi\right)
   -\sech\left(\sqrt{c_2}\xi\right)\right)}{-c_2\left[1-\left(\tanh\left(\sqrt{c_2}\xi\right)
   -\sech\left(\sqrt{c_2}\xi\right)\right)^2\right]}}$&$c_2>0,c_0>0$\\[0.25cm]
\noalign{\smallskip}\hline
\end{tabular}
\end{table*}
are equivalent to the following solutions
\begin{eqnarray*}
&F_8(\xi)=-{\frac {c_1}{2c_2}}+{\frac {\varepsilon\sqrt{\delta}}{2c_2}}\cosh\left(\sqrt{c_2}\xi\right),\delta>0,c_2>0,\\
&F_9(\xi)=-{\frac {c_1}{2c_2}}+{\frac {\varepsilon\sqrt{-\delta}}{2c_2}}\sinh\left(\sqrt{c_2}\xi\right),\delta<0,c_2>0,\\
&F_{10a}(\xi)=-{\frac {c_1}{2c_2}}+{\frac {\varepsilon\sqrt{\delta}}{2c_2}}\cos\left(\sqrt{-c_2}\xi\right),\delta>0,c_2<0,\\
&F_{10b}(\xi)=-{\frac {c_1}{2c_2}}+{\frac {\varepsilon\sqrt{\delta}}{2c_2}}\sin\left(\sqrt{-c_2}\xi\right),\delta>0,c_2<0.
\end{eqnarray*}
where $\delta=c_1^2-4c_0c_2$.\par
Let $A=\sqrt{\frac {c_0}{c_2}},B={\frac {c_1}{2c_2}}$,then $A^2-B^2=-{\frac {\delta}{4c_2^2}}$.
Thus,the condition $\delta>0$ gives that $A^2<B^2$,and the condition
$\delta<0$ implies that $A^2>B^2$.Hence,for $c_0>0,c_2>0$ we obtain from the
formulas (\ref{eq3:1}),the Lemma \ref{th3:3} and the Lemma \ref{th3:2} that
\begin{eqnarray*}
&Z_1(\xi)=-{\frac {c_1}{2c_2}}-\left(\sqrt{\frac {c_0}{c_2}}\sinh\left(\sqrt{c_2}\xi\right)
    -{\frac {c_1}{2c_2}}\cosh\left(\sqrt{c_2}\xi\right)\right),\\
&=\left\{\begin{array}{l}
-{\frac {c_1}{2c_2}}+{\frac {\varepsilon\sqrt{\delta}}{2c_2}}
\cosh\left(\sqrt{c_2}\xi-{\frac 12}\ln({\frac {c_1+2\sqrt{c_0c_2}}{c_1-2\sqrt{c_0c_2}}})\right)
\cong{F_8(\xi)},\delta>0,\\
-{\frac {c_1}{2c_2}}+{\frac {\varepsilon\sqrt{-\delta}}{2c_2}}
\sinh\left(\sqrt{c_2}\xi-{\frac 12}\ln({\frac {2\sqrt{c_0c_2}+c_1}{2\sqrt{c_0c_2}-c_1}})\right)
\cong{F_9(\xi)},\delta<0.
\end{array}\right.\\
&Z_2(\xi)=-{\frac {c_1}{2c_2}}+\sqrt{\frac {c_0}{c_2}}\sinh\left(\sqrt{c_2}\xi\right)
    -{\frac {c_1}{2c_2}}\cosh\left(\sqrt{c_2}\xi\right),\\
&=\left\{\begin{array}{l}
-{\frac {c_1}{2c_2}}+{\frac {\varepsilon\sqrt{\delta}}{2c_2}}
\cosh\left(\sqrt{c_2}\xi-{\frac 12}\ln({\frac {c_1+2\sqrt{c_0c_2}}{c_1-2\sqrt{c_0c_2}}})\right)
\cong{F_8(\xi)},\delta>0,\\
-{\frac {c_1}{2c_2}}+{\frac {\varepsilon\sqrt{-\delta}}{2c_2}}
\sinh\left(\sqrt{c_2}\xi-{\frac 12}\ln({\frac {2\sqrt{c_0c_2}+c_1}{2\sqrt{c_0c_2}-c_1}})\right)
\cong{F_9(\xi)},\delta<0.
\end{array}\right.
\end{eqnarray*}
Using the identities (\ref{eq3:2}) and the Lemma \ref{th3:4} yields
\begin{eqnarray*}
&Z_3(\xi)=-{\frac {c_1}{2c_2}}-\left({\frac {\sqrt{-c_0c_2}}{c_2}}\sin\left(\sqrt{-c_2}\xi\right)
  +{\frac {c_1}{2c_2}}\cos\left(\sqrt{-c_2}\xi\right)\right),\\
&\quad=\left\{\begin{array}{l}
-{\frac {c_1}{2c_2}}-{\frac {\sqrt{\delta}}{2c_2}}\cos(\sqrt{-c_2}\xi-\theta_1)
\cong{F_{10a}(\xi)|_{\varepsilon=-1}},\delta>0,c_0>0,c_2<0,\\
-{\frac {c_1}{2c_2}}-{\frac {\sqrt{\delta}}{2c_2}}\sin(\sqrt{-c_2}\xi+\theta_2)
\cong{F_{10b}(\xi)|_{\varepsilon=-1}},\delta>0,c_0>0,c_2<0.
\end{array}\right.\\
&Z_4(\xi)=-{\frac {c_1}{2c_2}}+\left({\frac {\sqrt{-c_0c_2}}{c_2}}\sin\left(\sqrt{-c_2}\xi\right)
  +{\frac {c_1}{2c_2}}\cos\left(\sqrt{-c_2}\xi\right)\right),\\
&\quad=\left\{\begin{array}{l}
-{\frac {c_1}{2c_2}}-{\frac {\sqrt{\delta}}{2c_2}}\cos(\sqrt{-c_2}\xi-\theta_1)
\cong{F_{10a}(\xi)|_{\varepsilon=-1}},\delta>0,c_0>0,c_2<0,\\
-{\frac {c_1}{2c_2}}-{\frac {\sqrt{\delta}}{2c_2}}\sin(\sqrt{-c_2}\xi+\theta_2)
\cong{F_{10b}(\xi)|_{\varepsilon=-1}},\delta>0,c_0>0,c_2<0,\\
\end{array}\right.
\end{eqnarray*}
where $\theta_1=\arctan\left({\frac {2\sqrt{-c_0c_2}}{c_1}}\right),
\theta_2=\arctan\left({\frac {c_1}{2\sqrt{-c_0c_2}}}\right).$
\newline
Set $A=1-c_1^2+4c_0c_4,B=1+c_1^2-4c_0c_4$ then we find
$A^2-B^2=-4\delta,A-B=-2\delta$.Therefore,from the condition
$\delta>0$ we obtain that $A^2<B^2$ and $A<B$,and the condition $\delta<0$ implies
that $A^2>B^2$ and $A>B$.Hence,it follows from the Lemma \ref{th3:3} and
the Lemma \ref{th3:2} that
\begin{eqnarray*}
&Z_5(\xi)=-{\frac {c_1}{2c_2}}+{\frac 1{4c_2}}
\left[(1-c_1^2+4c_0c_4)\sinh\left(\sqrt{c_2}\xi\right)
 -(1+c_1^2-4c_0c_4)\cosh\left(\sqrt{c_2}\xi\right)\right],\\
&\quad=\left\{\begin{array}{l}
-{\frac {c_1}{2c_2}}-{\frac {\sqrt{\delta}}{2c_2}}\cosh\left(\sqrt{c_2}\xi-{\frac 12}\ln({\frac 1\delta})\right)
\cong{F_{8}(\xi)|_{\varepsilon=-1}},\delta>0,c_2>0,\\
-{\frac {c_1}{2c_2}}+{\frac {\sqrt{-\delta}}{2c_2}}\sinh\left(\sqrt{c_2}\xi-{\frac 12}\ln({\frac {-1}\delta})\right)\cong{F_9(\xi)|_{\varepsilon=1}},\delta<0,c_2>0.
\end{array}\right.
\end{eqnarray*}
Using $\sinh{x}+\cosh{x}=e^x$ and the Lemma \ref{th3:1},we obtain
\begin{eqnarray*}
&Z_6(\xi)=-{\frac {c_1}{2c_2}}+{\frac 1{8c_2^2}}\left[4c_2^2e^{\sqrt{c_2}\xi}+(c_1^2-4c_0c_4)e^{-\sqrt{c_2}\xi}\right],\\
&\quad=\left\{\begin{array}{l}
-{\frac {c_1}{2c_2}}+{\frac {\sqrt{\delta}}{2c_2}}\cosh\left(\sqrt{c_2}\xi+{\frac 12}\ln({\frac {4c_2^2}{\delta}})\right)
\cong{F_{8}(\xi)|_{\varepsilon=1}},\delta>0,c_2>0,\\
-{\frac {c_1}{2c_2}}+{\frac {\sqrt{-\delta}}{2c_2}}\sinh\left(\sqrt{c_2}\xi+{\frac 12}\ln({\frac {4c_2^2}{-\delta}})\right)\cong{F_9(\xi)|_{\varepsilon=1}},\delta<0,c_2>0.
\end{array}\right.
\end{eqnarray*}
The following identities are useful to prove the equivalence
relations for solutions of the auxiliary equations and we can
prove these identities by means of the trigonometric and
hyperbolic function identities with the aid of some direct calculations.
\begin{eqnarray}
\label{eq3:3}
&\tan{\eta}+\sec{\eta}=\tan\left({\frac {\eta}{2}}+{\frac {\pi}4}\right),
\tan{\eta}-\sec{\eta}=-\cot\left({\frac {\eta}{2}}+{\frac {\pi}4}\right),\\
\label{eq3:4}
&\cot{\eta}+\csc{\eta}=\cot\left({\frac {\eta}2}\right),\cot{\eta}-\csc{\eta}=-\tan\left({\frac {\eta}2}\right)\\
\label{eq3:5}
&\tanh{\eta}+i\sech{\eta}=\tanh\left({\frac {\eta}2}+{\frac {\pi{i}}{4}}\right),
\tanh{\eta}-i\sech{\eta}=\coth\left({\frac {\eta}2}+{\frac {\pi{i}}{4}}\right).
\end{eqnarray}
Now the identities (\ref{eq3:3}),(\ref{eq3:4}),(\ref{eq3:5}) and the trigonometric
identities are employed to obtain
\begin{eqnarray*}
\begin{array}{l}
Z_7(\xi)=-{\frac {c_1+2\sqrt{-c_0c_2}\tan\left({\frac {\sqrt{-c_2}}{2}}\xi+{\frac {\pi}{4}}\right)}{c_2\left(1+\tan^2\left({\frac {\sqrt{-c_2}}{2}}\xi+{\frac {\pi}{4}}\right)\right)}}=-{\frac {c_1+2\sqrt{-c_0c_2}\tan\left({\frac {\sqrt{-c_2}}{2}}\xi+{\frac {\pi}{4}}\right)}{c_2\sec^2\left({\frac {\sqrt{-c_2}}{2}}\xi+{\frac {\pi}{4}}\right)}},\\
\quad=-{\frac 1{c_2}}\left[c_1+2\sqrt{-c_0c_2}\tan\left({\frac {\sqrt{-c_2}}{2}}\xi+{\frac {\pi}{4}}\right)\right]\cos^2\left({\frac {\sqrt{-c_2}}{2}}\xi+{\frac {\pi}{4}}\right)\cong{Z_3(\xi)},\\
Z_8(\xi)=-{\frac {c_1+2\sqrt{-c_0c_2}\cot\left({\frac {\sqrt{-c_2}}{2}}\xi+{\frac {\pi}{4}}\right)}{c_2\left(1+\cot^2\left({\frac {\sqrt{-c_2}}{2}}\xi+{\frac {\pi}{4}}\right)\right)}}
=-{\frac {c_1-2\sqrt{-c_0c_2}\cot\left({\frac {\sqrt{-c_2}}{2}}\xi+{\frac {\pi}{4}}\right)}{c_2\csc^2\left({\frac {\sqrt{-c_2}}{2}}\xi+{\frac {\pi}{4}}\right)}},\\
\quad=-{\frac 1{c_2}}\left[c_1-2\sqrt{-c_0c_2}\cot\left({\frac {\sqrt{-c_2}}{2}}\xi+{\frac {\pi}{4}}\right)\right]sin^2\left({\frac {\sqrt{-c_2}}{2}}\xi+{\frac {\pi}{4}}\right)\cong{Z_4(\xi)},\\
Z_9(\xi)={\frac {-c_1+2\sqrt{-c_0c_2}\cot\left({\frac {\sqrt{-c_2}}{2}}\xi\right)}{c_2\left(1+\cot^2\left({\frac {\sqrt{-c_2}}{2}}\xi\right)\right)}}
={\frac {-c_1+2\sqrt{-c_0c_2}\cot\left({\frac {\sqrt{-c_2}}{2}}\xi\right)}{c_2\csc^2\left({\frac {\sqrt{-c_2}}{2}}\xi\right)}},\\
\quad=-{\frac 1{c_2}}\left[c_1-2\sqrt{-c_0c_2}\cot\left({\frac {\sqrt{-c_2}}{2}}\xi\right)\right]\sin^2\left({\frac {\sqrt{-c_2}}{2}}\xi\right)={Z_4(\xi)},\\
Z_{10}(\xi)={\frac {-c_1-2\sqrt{-c_0c_2}\tan\left({\frac {\sqrt{-c_2}}{2}}\xi\right)}{c_2\left(1+\tan^2\left({\frac {\sqrt{-c_2}}{2}}\xi\right)\right)}}
={\frac {-c_1-2\sqrt{-c_0c_2}\tan\left({\frac {\sqrt{-c_2}}{2}}\xi\right)}{c_2\sec^2\left({\frac {\sqrt{-c_2}}{2}}\xi\right)}},\\
\quad=-{\frac 1{c_2}}\left[c_1+2\sqrt{-c_0c_2}\tan\left({\frac {\sqrt{-c_2}}{2}}\xi\right)\right]\cos^2\left({\frac {\sqrt{-c_2}}{2}}\xi\right)={Z_3(\xi)},\\
Z_{11}(\xi)=-{\frac {c_1-2\sqrt{c_0c_2}\tanh\left({\frac {\sqrt{c_2}}{2}}\xi+{\frac {\pi{i}}{4}}\right)}{c_2\left(1-\tanh^2\left({\frac {\sqrt{c_2}}{2}}\xi+{\frac {\pi{i}}{4}}\right)\right)}}
=-{\frac {c_1-2\sqrt{c_0c_2}\tanh\left({\frac {\sqrt{c_2}}{2}}\xi+{\frac {\pi{i}}{4}}\right)}{c_2\sech^2\left({\frac {\sqrt{c_2}}{2}}\xi+{\frac {\pi{i}}{4}}\right)}},\\
\quad=-{\frac 1{c_2}}\left[c_1-2\sqrt{c_0c_2}\tanh\left({\frac {\sqrt{c_2}}{2}}\xi+{\frac {\pi{i}}{4}}\right)\right]\cosh^2\left({\frac {\sqrt{c_2}}{2}}\xi+{\frac {\pi{i}}{4}}\right)\cong{Z_2(\xi)},\\
Z_{12}(\xi)=-{\frac {c_1-2\sqrt{c_0c_2}\coth\left({\frac {\sqrt{c_2}}{2}}\xi+{\frac {\pi{i}}{4}}\right)}{c_2\left(1-\coth^2\left({\frac {\sqrt{c_2}}{2}}\xi+{\frac {\pi{i}}{4}}\right)\right)}}
=-{\frac {c_1-2\sqrt{c_0c_2}\coth\left({\frac {\sqrt{c_2}}{2}}\xi+{\frac {\pi{i}}{4}}\right)}{c_2\csch^2\left({\frac {\sqrt{c_2}}{2}}\xi+{\frac {\pi{i}}{4}}\right)}},\\
\quad=-{\frac 1{c_2}}\left[c_1-2\sqrt{c_0c_2}\coth\left({\frac {\sqrt{c_2}}{2}}\xi+{\frac {\pi{i}}{4}}\right)\right]\sinh^2\left({\frac {\sqrt{c_2}}{2}}\xi+{\frac {\pi{i}}{4}}\right)\cong{Z_1(\xi)}.
\end{array}
\end{eqnarray*}
In accordance with the transitivity,we have proved the equivalence relations $Z_i(\xi)\cong{F_{10a}(\xi)|_{\varepsilon=-1}}\\
(c_0>0,c_2<0,\delta>0,i=7,8,9,10)$,
$Z_{i}(\xi)\cong{F_8(\xi)}~(c_0>0,c_2>0,\delta>0,i=11,12)$ and $Z_{i}(\xi)\cong{F_9(\xi)}~(c_0>0,c_2>0,\delta<0,i=11,12)$.\par
For the following two solutions~\cite{RefB-1} of Eq.(\ref{eq1:7})
\begin{eqnarray*}
\begin{array}{l}
y_1(\xi)={\frac {-c_1+\sqrt{\delta}}{2c_2}}+{\frac {\sqrt{\delta}}{c_2}}\sinh^2\left({\frac {\sqrt{c_2}}{2}}\xi\right),\delta=c_1^2-4c_0c_2>0,c_2>0,\\
y_2(\xi)={\frac {-c_1-\sqrt{\delta}}{2c_2}}-{\frac {\sqrt{\delta}}{c_2}}\sinh^2\left({\frac {\sqrt{c_2}}{2}}\xi\right),\delta=c_1^2-4c_0c_2>0,c_2>0,
\end{array}
\end{eqnarray*}
we have $y_1(\xi)=F_8(\xi)|_{\varepsilon=1}$ and $y_2(\xi)=F_8(\xi)|_{\varepsilon=-1}$
which can be obtained by substituting the identity $\sinh^2\left({\frac {\sqrt{c_2}}{2}}\xi\right)={\frac 12}\left(\cosh\left(\sqrt{c_2}\xi\right)-1\right)$
into $y_1(\xi)$ and $y_2(\xi)$.
\section{The classification of the solutions}\label{sec:4}
The classification of the solutions of Eq.(\ref{eq1:1}) is very important
to clarify whether the solutions given in literatures are new or
equivalent in the sense of wave translation.If this problem is clear then
we can choose the new solutions and avoid the repeated solutions effectively.
Based on the equivalence relations of the solutions for Eq.(\ref{eq1:1})
as proved in Sec.\ref{sec:3},by removing all those repeated solutions and
collecting the simple independent solutions presented in the literatures,
we now conclude that the solutions of Eq.(\ref{eq1:1}) can be classified
into six classes such as the hyperbolic function solutions,the trigonometric
function solutions,the elliptic function solutions,the exponential function
solutions,the polynomial solutions and the rational solutions.In other words,
the six classes of solutions containing thirty eight independent solutions can be
divided into the five cases as below.
\begin{cas}
\label{case-1}
$c_0=c_1=0$.
\begin{eqnarray*}
\begin{array}{l}
F_1(\xi)={\frac {2c_2}{\varepsilon\sqrt{\Delta}\cosh\left(\sqrt{c_2}\xi\right)-c_3}},\Delta>0,c_2>0,\\
F_2(\xi)={\frac {2c_2}{\varepsilon\sqrt{-\Delta}\sinh\left(\sqrt{c_2}\xi\right)-c_3}},\Delta<0,c_2>0,\\
F_{3a}(\xi)={\frac {2c_2}{\varepsilon\sqrt{\Delta}\cos\left(\sqrt{-c_2}\xi\right)-c_3}},\\
\quad\,F_{3b}(\xi)={\frac {2c_2}{\varepsilon\sqrt{\Delta}\sin\left(\sqrt{-c_2}\xi\right)-c_3}},\Delta>0,c_2<0,\\
F_4(\xi)=-{\frac {c_2}{c_3}}\left[1+\varepsilon\tanh\left({\frac {\sqrt{c_2}}2}\xi\right)\right],\Delta=0,c_2>0,\\
F_5(\xi)=-{\frac {c_2}{c_3}}\left[1+\varepsilon\coth\left({\frac {\sqrt{c_2}}2}\xi\right)\right],\Delta=0,c_2>0,\\
F_6(\xi)={\frac {\varepsilon}{\sqrt{c_4}\xi}},c_2=c_3=0,c_4>0,\\
F_7(\xi)={\frac {4c_3}{c_3^2\xi^2-4c_4}},c_2=0,
\end{array}
\end{eqnarray*}
where $\Delta=c_3^2-4c_2c_4,\varepsilon=\pm{1}$.
\end{cas}
\begin{cas}
\label{case-2}
$c_3=c_4=0$.
\begin{eqnarray*}
\begin{array}{l}
F_8(\xi)=-{\frac {c_1}{2c_2}}+{\frac {\varepsilon\sqrt{\delta}}{2c_2}}\cosh\left(\sqrt{c_2}\xi\right),\delta>0,c_2>0,\\
F_9(\xi)=-{\frac {c_1}{2c_2}}+{\frac {\varepsilon\sqrt{-\delta}}{2c_2}}\sinh\left(\sqrt{c_2}\xi\right),\delta<0,c_2>0,\\
F_{10a}(\xi)=-{\frac {c_1}{2c_2}}+{\frac {\varepsilon\sqrt{\delta}}{2c_2}}\cos\left(\sqrt{-c_2}\xi\right),\\
\quad\,F_{10b}(\xi)=-{\frac {c_1}{2c_2}}+{\frac {\varepsilon\sqrt{\delta}}{2c_2}}\sin\left(\sqrt{-c_2}\xi\right),\delta>0,c_2<0,\\
F_{11}(\xi)=-{\frac {c_1}{2c_2}}+e^{\varepsilon\sqrt{c_2}\xi},\delta=0,c_2>0,\\
F_{12}(\xi)=\varepsilon\sqrt{c_0}\xi,c_1=c_2=0,\\
F_{13}(\xi)=-{\frac {c_0}{c_1}}+{\frac {c_1}4}\xi^2,c_2=0,
\end{array}
\end{eqnarray*}
where $\delta=c_1^2-4c_0c_2,\varepsilon=\pm{1}$.
\end{cas}
\begin{cas}
\label{cse-3}
$c_1=c_3=0$.
\begin{eqnarray*}
\begin{array}{l}
F_{14}(\xi)=\varepsilon\sqrt{-\frac {c_2}{2c_4}}\tanh\left(\sqrt{-\frac {c_2}2}\xi\right),\Delta_1=0,c_2<0,c_4>0,\\
F_{15}(\xi)=\varepsilon\sqrt{-\frac {c_2}{2c_4}}\coth\left(\sqrt{-\frac {c_2}2}\xi\right),\Delta_1=0,c_2<0,c_4>0,\\
F_{16a}(\xi)=\varepsilon\sqrt{\frac {c_2}{2c_4}}\tan\left(\sqrt{\frac {c_2}2}\xi\right),\\
\quad\,F_{16b}(\xi)=\varepsilon\sqrt{\frac {c_2}{2c_4}}\cot\left(\sqrt{\frac {c_2}2}\xi\right),\Delta_1=0,c_2>0,c_4>0,\\
F_{17}(\xi)=\sqrt{\frac {-c_2m^2}{c_4(m^2+1)}}\sn\left(\sqrt{\frac {-c_2}{m^2+1}}\xi\right), c_0={\frac {c_2^2m^2}{c_4(m^2+1)^2}},c_2<0,c_4>0,\\
F_{18}(\xi)=\sqrt{\frac {-c_2m^2}{c_4(2m^2-1)}}\cn\left(\sqrt{\frac {c_2}{2m^2-1}}\xi\right), c_0={\frac {c_2^2m^2(m^2-1)}{c_4(2m^2-1)^2}},c_2>0,c_4<0,\\
F_{19}(\xi)=\sqrt{\frac {-c_2}{c_4(2-m^2)}}\dn\left(\sqrt{\frac {c_2}{2-m^2}}\xi\right), c_0={\frac {c_2^2(1-m^2)}{c_4(2-m^2)^2}},c_2>0,c_4<0,\\
F_{20}(\xi)=\varepsilon\left(-{\frac {4c_0}{c_4}}\right)^{\frac 14}\ds\left((-4c_0c_4)^{\frac 14}\xi,{\frac {\sqrt{2}}2}\right),c_2=0,c_0c_4<0,\\
F_{21}(\xi)=\varepsilon\left({\frac {c_0}{c_4}}\right)^{\frac 14}\left[\ns\left(2(c_0c_4)^{\frac 14}\xi,{\frac {\sqrt{2}}2}\right)
     +\cs\left(2(c_0c_4)^{\frac 14}\xi,{\frac {\sqrt{2}}2}\right)\right],c_2=0,c_0c_4>0,
\end{array}
\end{eqnarray*}
where $\Delta_1=c_2^2-4c_0c_4,\varepsilon=\pm{1}$.
\end{cas}
\begin{cas}
\label{cse-4}
$c_2=c_4=0$.
\begin{eqnarray*}
F_{22}(\xi)=\wp\left({\frac {\sqrt{c_3}}2}\xi,g_2,g_3\right),g_2=-{\frac {4c_1}{c_3}},g_3=-{\frac {4c_0}{c_3}},c_3>0.
\end{eqnarray*}
\end{cas}
\begin{cas}
\label{case-5}
$c_0=0$.
\begin{eqnarray*}
\begin{array}{l}
F_{23}(\xi)=-{\frac {8c_2\tanh^2\left(\sqrt{-\frac {c_2}{12}}\xi\right)}
   {3c_3\left(3+\tanh^2\left(\sqrt{-\frac {c_2}{12}}\xi\right)\right)}},c_2<0,c_1={\frac {8c_2^2}{27c_3}},c_4={\frac {c_3^2}{4c_2}},\\
F_{24}(\xi)=-{\frac {8c_2\coth^2\left(\sqrt{-\frac {c_2}{12}}\xi\right)}
   {3c_3\left(3+\coth^2\left(\sqrt{-\frac {c_2}{12}}\xi\right)\right)}},c_2<0,c_1={\frac {8c_2^2}{27c_3}},c_4={\frac {c_3^2}{4c_2}},\\
F_{25}(\xi)={\frac {8c_2\tan^2\left(\sqrt{\frac {c_2}{12}}\xi\right)}
   {3c_3\left(3-\tan^2\left(\sqrt{\frac {c_2}{12}}\xi\right)\right)}},c_2>0,c_1={\frac {8c_2^2}{27c_3}},c_4={\frac {c_3^2}{4c_2}},\\
F_{26}(\xi)={\frac {8c_2\cot^2\left(\sqrt{\frac {c_2}{12}}\xi\right)}
   {3c_3\left(3-\cot^2\left(\sqrt{\frac {c_2}{12}}\xi\right)\right)}},c_2>0,c_1={\frac {8c_2^2}{27c_3}},c_4={\frac {c_3^2}{4c_2}},\\
F_{27}(\xi)=-{\frac {c_3}{4c_4}}\left[1+\varepsilon{\sn}\left({\frac {c_3}
   {4m\sqrt{c_4}}}\xi\right)\right],c_4>0,c_1={\frac {c_3^3(m^2-1)}{32m^2c_4^2}},c_2={\frac {c_3^2(5m^2-1)}{16m^2c_4}},\\
F_{28}(\xi)=-{\frac {c_3}{4c_4}}\left[1+{\frac \varepsilon{m{\sn}\left({\frac {c_3}
   {4m\sqrt{c_4}}}\xi\right)}}\right],c_4>0,c_1={\frac {c_3^3(m^2-1)}{32m^2c_4^2}},c_2={\frac {c_3^2(5m^2-1)}{16m^2c_4}},\\
F_{29}(\xi)=-{\frac {c_3}{4c_4}}\left[1+\varepsilon{m\sn}\left({\frac {c_3}
   {4\sqrt{c_4}}}\xi\right)\right],c_4>0,c_1={\frac {c_3^3(1-m^2)}{32c_4^2}},c_2={\frac {c_3^2(5-m^2)}{16c_4}},\\
F_{30}(\xi)=-{\frac {c_3}{4c_4}}\left[1+{\frac \varepsilon{{\sn}\left({\frac {c_3}
   {4\sqrt{c_4}}}\xi\right)}}\right],c_4>0,c_1={\frac {c_3^3(1-m^2)}{32c_4^2}},c_2={\frac {c_3^2(5-m^2)}{16c_4}},\\
F_{31}(\xi)=-{\frac {c_3}{4c_4}}\left[1+\varepsilon{\cn}\left({-\frac {c_3}
   {4m\sqrt{-c_4}}}\xi\right)\right],c_4<0,c_1={\frac {c_3^3}{32m^2c_4^2}},c_2={\frac {c_3^2(4m^2+1)}{16m^2c_4}},\\
F_{32}(\xi)=-{\frac {c_3}{4c_4}}\left[1+{\frac {\varepsilon\sqrt{1-m^2}{\sn}\left({-\frac {c_3}
   {4m\sqrt{-c_4}}}\xi\right)}{{\dn}\left({-\frac {c_3}{4m\sqrt{-c_4}}}\xi\right)}}\right],c_4<0,c_1={\frac {c_3^3}{32m^2c_4^2}},c_2={\frac {c_3^2(4m^2+1)}{16m^2c_4}},\\
F_{33}(\xi)=-{\frac {c_3}{4c_4}}\left[1+{\frac {\varepsilon}{\sqrt{1-m^2}}}{\dn}
\left({\frac {c_3}{4\sqrt{c_4(m^2-1)}}}\xi\right)\right],\\
\qquad\,c_4<0,c_1={\frac {c_3^3m^2}{32c_4^2(m^2-1)}},c_2={\frac {c_3^2(5m^2-4)}{16c_4(m^2-1)}},\\
F_{34}(\xi)=-{\frac {c_3}{4c_4}}\left[1+{\frac {\varepsilon}{{\dn}\left({\frac {c_3}
   {4\sqrt{c_4(m^2-1)}}}\xi\right)}}\right],c_4<0,c_1={\frac {c_3^3m^2}{32c_4^2(m^2-1)}},c_2={\frac {c_3^2(5m^2-4)}{16c_4(m^2-1)}},\\
F_{35}(\xi)=-{\frac {c_3}{4c_4}}\left[1+{\frac {\varepsilon}{{\cn}\left({\frac {c_3}
   {4\sqrt{c_4(1-m^2)}}}\xi\right)}}\right],c_4>0,c_1={\frac {c_3^3}{32c_4^2(1-m^2)}},c_2={\frac {c_3^2(4m^2-5)}{16c_4(m^2-1)}},\\
F_{36}(\xi)=-{\frac {c_3}{4c_4}}\left[1+{\frac {\varepsilon{\dn}\left({\frac {c_3}
   {4\sqrt{c_4(1-m^2)}}}\xi\right)}{\sqrt{1-m^2}{\cn}\left({\frac {c_3}{4\sqrt{c_4(1-m^2)}}}\xi\right)}}\right],\\
\qquad\,c_4>0,c_1={\frac {c_3^3}{32c_4^2(1-m^2)}},c_2={\frac {c_3^2(4m^2-5)}{16c_4(m^2-1)}},\\
F_{37}(\xi)=-{\frac {c_3}{4c_4}}\left[1+\varepsilon{\dn}\left({-\frac {c_3}
   {4\sqrt{-c_4}}}\xi\right)\right],c_4<0,c_1={\frac {c_3^3m^2}{32c_4^2}},c_2={\frac {c_3^2(m^2+4)}{16c_4}},\\
F_{38}(\xi)=-{\frac {c_3}{4c_4}}\left[1+{\frac {\varepsilon\sqrt{1-m^2}}{{\dn}\left({-\frac {c_3}
   {4\sqrt{-c_4}}}\xi\right)}}\right],c_4<0,c_1={\frac {c_3^3m^2}{32c_4^2}},c_2={\frac {c_3^2(m^2+4)}{16c_4}}
\end{array}
\end{eqnarray*}
where $\varepsilon=\pm{1}$.
\end{cas}
\section{New traveling wave solutions for mCH equation}
\label{sec:5}
\setcounter{cas}{0}
As an illustrative example below we consider the modified
Camassa--Holm (mCH) equation in the form
\begin{equation}
\label{eq5:1}
u_t-u_{xxt}+3u^2u_x=2u_{x}u_{xx}+uu_{xxx},
\end{equation}
was due to Wazwaz~\cite{RefJ-35} obtained by using the term $u^2u_x$
instead of the nonlinear convection term $uu_x$ in Camassa-Holm equation.
Wazwaz~\cite{RefJ-35,RefJ-36} also found some exact traveling wave solutions
of wave speed $c=1/2,1,2$ for Eq.(\ref{eq5:1}).Some traveling wave solutions
of wave speed $c=1/3$ and some peakon solutions of wave speed $c=3$ for
Eq.(\ref{eq5:1}) were obtained by Wang and Tang in~\cite{RefJ-37}.
A rational solution and more exact traveling wave solutions of Eq.(\ref{eq5:1})
were also constructed by Liang and Jeffrey in~\cite{RefJ-38}.\par
In the present paper we are interested in finding more new exact traveling wave
solutions for Eq.(\ref{eq5:1}) through use of the general elliptic equation
expansion method equipped with our new classified solutions in Sec.\ref{sec:4}.
In order that we make the wave transformation $u(x,t)=u(\xi),\xi=x-{\omega}t$
and change the Eq.(\ref{eq5:1}) into the form
\begin{equation}
\label{eq5:2}
-\omega{u^\prime}+\omega{u^{\prime\prime\prime}}+3u^2u^\prime
-2{u^\prime}u^{\prime\prime}-uu^{\prime\prime\prime}=0.
\end{equation}
According to the balancing principle to balance $uu^{\prime\prime\prime}$
with $u^2{u^\prime}$ yields $m=2$, therefore the Eq.(\ref{eq5:2}) admits
the solution in the form
\begin{equation}
\label{eq5:3}
u(\xi)=a_0+a_1F(\xi)+a_2F^2(\xi),
\end{equation}
where $F(\xi)$ expresses the solution of Eq.(\ref{eq1:1}) and
$a_k~(k=0,1,2)$ are constants to be determined.\par
Substituting (\ref{eq5:3}) into (\ref{eq5:2}) along with Eq.(\ref{eq1:1}) and setting
the coefficients of $F^jF^\prime~(j=0,1,2,3,4,5)$ to zero,we obtain
the following set of algebraic equations
\begin{equation}
\label{eq5:4}
\left\{\begin{array}{l}
6a_2^3-48a_2^2c_4=0,\\
15a_1a_2^2-50a_1a_2c_4-35a_2^2c_3=0,\\ 12a_0a_2^2-24a_0a_2c_4+12a_1^2a_2-10a_1^2c_4-34a_1a_2c_3-24a_2^2c_2+24a_2c_4\omega=0,\\
3a_0^2a_1-a_0a_1c_2-3a_0a_2c_1-a_1^2c_1-4a_1a_2c_0+a_1c_2\omega+3a_2c_1\omega-a_1\omega=0,\\ 18a_0a_1a_2-6a_0a_1c_4-15a_0a_2c_3+3a_1^3-6a_1^2c_3-21a_1a_2c_2+6a_1c_4\omega\\
\quad-15a_2^2c_1+15a_2c_3\omega=0,\\ 6a_0^2a_2+6a_0a_1^2-3a_0a_1c_3-8a_0a_2c_2-3a_1^2c_2-2a_2\omega-11a_1a_2c_1+3a_1c_3\omega\\
\quad-8a_2^2c_0+8a_2c_2\omega=0.
\end{array}\right.
\end{equation}
{\bf A.} When $c_0=c_1=0$,the Eq.(\ref{eq5:4}) solves that
\begin{eqnarray}
\label{eq5:5}
&a_0={\frac 13}\left(-\omega+\sqrt{2\omega(3-\omega)}\right),a_1=0,a_2=8c_4,c_2={\frac 18}\sqrt{2\omega(3-\omega)},c_3=0,\\
\label{eq5:6}
&a_0={\frac 13}\left(-\omega-\sqrt{2\omega(3-\omega)}\right),a_1=0,a_2=8c_4,c_2=-{\frac 18}\sqrt{2\omega(3-\omega)},c_3=0,\\
\label{eq5:7}
&a_0={\frac 13}\left(-\omega+\sqrt{2\omega(3-\omega)}\right),a_1=2c_3,a_2=0,c_2={\frac 12}\sqrt{2\omega(3-\omega)},c_4=0,\\
\label{eq5:8}
&a_0={\frac 13}\left(-\omega-\sqrt{2\omega(3-\omega)}\right),a_1=2c_3,a_2=0,c_2=-{\frac 12}\sqrt{2\omega(3-\omega)},c_4=0,\\
\label{eq5:9}
&a_0={\frac 13}\left(-\omega+\sqrt{2\omega(3-\omega)}\right),a_1=\pm\left(8c_4\right)^{\frac 12}\left(2\omega(3-\omega)\right)^{\frac 14},
a_2=8c_4,\nonumber\\
&{\quad}c_2={\frac 12}\sqrt{2\omega(3-\omega)},c_4=\pm{\frac 12}\left(8c_4\right)^{\frac 12}\left(2\omega(3-\omega)\right)^{\frac 14},\\
\label{eq5:10}
&a_0={\frac 13}\left(-\omega+\sqrt{2\omega(3-\omega)}\right),a_1=2c_3,c_2={\frac 12}\sqrt{2\omega(3-\omega)},c_4=0,\\
\label{eq5:11}
&a_0={\frac 13}\left(-\omega-\sqrt{2\omega(3-\omega)}\right),a_1=2c_3,c_2=-{\frac 12}\sqrt{2\omega(3-\omega)},c_4=0.
\end{eqnarray}
Inserting (\ref{eq5:5})--(\ref{eq5:11}) together with the solutions
given by Case~\ref{case-1} into (\ref{eq5:3}),we obtain the exact
traveling wave solutions of Eq.(\ref{eq5:1}) as follows
\begin{eqnarray*}
\begin{array}{l}
u_{1a}(x,t)=-{\frac {\omega}3}+{\frac 13}\sqrt{2\omega(3-\omega)}-\sqrt{2\omega(3-\omega)}\sech^2\eta,\\
u_{1b}(x,t)=-{\frac {\omega}3}+{\frac 13}\sqrt{2\omega(3-\omega)}+\sqrt{2\omega(3-\omega)}\csch^2\eta,\\
u_{2a}(x,t)=-{\frac {\omega}3}-{\frac 13}\sqrt{2\omega(3-\omega)}+\sqrt{2\omega(3-\omega)}\sec^2\eta,\\
u_{2b}(x,t)=-{\frac {\omega}3}-{\frac 13}\sqrt{2\omega(3-\omega)}+\sqrt{2\omega(3-\omega)}\csc^2\eta,\\
u_3(x,t)=-{\frac {\omega}3}+{\frac 13}\sqrt{2\omega(3-\omega)}
+{\frac {2\sqrt{2\omega(3-\omega)}}{\varepsilon\cosh\eta-1}},\\
u_{4a}(x,t)=-{\frac {\omega}3}-{\frac 13}\sqrt{2\omega(3-\omega)}
  -{\frac {2\sqrt{2\omega(3-\omega)}}{\varepsilon\cos\eta-1}},\\
u_{4b}(x,t)=-{\frac {\omega}3}-{\frac 13}\sqrt{2\omega(3-\omega)}
  -{\frac {2\sqrt{2\omega(3-\omega)}}{\varepsilon\sin\eta-1}},\\
u_{5a}(x,t)=-{\frac {\omega}3}-{\frac 23}\sqrt{2\omega(3-\omega)}+\sqrt{2\omega(3-\omega)}\tanh^2\eta,\\
u_{5b}(x,t)=-{\frac {\omega}3}-{\frac 23}\sqrt{2\omega(3-\omega)}+\sqrt{2\omega(3-\omega)}\coth^2\eta,\\
u_{6a}(x,t)=u_{1a}(x,t),u_{6b}(x,t)=u_{1b}(x,t),c_4=0,\\
u_{7a}(x,t)=u_{2a}(x,t),u_{7b}(x,t)=u_{2b}(x,t),c_4=0,\\
\qquad\eta={\frac 12}\left({\frac {\omega(3-\omega)}2}\right)^{\frac 14}(x-{\omega}t),
0<\omega<3.
\end{array}
\end{eqnarray*}
\newline
{\bf B.} When $c_3=c_4=0$,we cannot find the solutions of Eq.(\ref{eq5:4}) so
we can not obtain the exact traveling wave solutions for Eq.(\ref{eq5:1}).
\newline
{\bf C.} When $c_1=c_3=0$,we have
\newline
(1)\,For $c_0={\frac {c_2^2m^2}{c_4(m^2+1)^2}},c_2<0,c_4>0$,
the Eq.(\ref{eq5:4}) solves that
\begin{eqnarray*}
&a_0=-{\frac {\omega}3}+{\frac {m^2+1}3}\left({\frac {2\omega(3-\omega)}{m^4-m^2+1}}\right)^{\frac 12},a_1=0,\\
&{\quad}a_2=8c_4,c_2=-{\frac {m^2+1}8}\left({\frac {2\omega(3-\omega)}{m^4-m^2+1}}\right)^{\frac 12},
\end{eqnarray*}
which leads the Jacobian snoidal wave solution of Eq.(\ref{eq5:1}) as
\begin{eqnarray}
\label{eq5:12}
&u(x,t)=-{\frac {\omega}3}-{\frac {m^2+1}3}\left({\frac {2\omega(3-\omega)}{m^4-m^2+1}}\right)^{\frac 12}+m^2\left({\frac {2\omega(3-\omega)}{m^4-m^2+1}}\right)^{\frac 12}\sn^2(\eta,m),\nonumber\\
&\quad\eta={\frac 12}\left({\frac {\omega(3-\omega)}{2(m^4-m^2+1)}}\right)^{\frac 14}(x-{\omega}t),0<\omega<3.
\end{eqnarray}
\par\noindent
(2)\,For $c_0={\frac {c_2^2m^2(m^2-1)}{c_4(2m^2-1)^2}},c_2>0,c_4<0$,
the solution of Eq.(\ref{eq5:4}) is found to be
\begin{eqnarray*}
&a_0=-{\frac {\omega}3}+{\frac {2m^2-1}3}\left({\frac {2\omega(3-\omega)}{m^4-m^2+1}}\right)^{\frac 12},a_1=0,\\
&{\quad}a_2=8c_4,c_2={\frac {2m^2-1}3}\left({\frac {2\omega(3-\omega)}{m^4-m^2+1}}\right)^{\frac 12},
\end{eqnarray*}
which gives the Jacobian cnoidal wave solution of Eq.(\ref{eq5:1}) as following
\begin{eqnarray}
\label{eq5:13}
&u(x,t)=-{\frac {\omega}3}+{\frac {2m^2-1}3}\left({\frac {2\omega(3-\omega)}{m^4-m^2+1}}\right)^{\frac 12}
-m^2\left({\frac {2\omega(3-\omega)}{m^4-m^2+1}}\right)^{\frac 12}\cn^2(\eta,m),\nonumber\\
&\quad\eta={\frac 12}\left({\frac {\omega(3-\omega)}{2(m^4-m^2+1)}}\right)^{\frac 14}(x-{\omega}t),0<\omega<3.
\end{eqnarray}
\par\noindent
(3)\,For $c_0={\frac {c_2^2(1-m^2)}{c_4(2-m^2)^2}},c_2>0,c_4<0$,
the solution of Eq.(\ref{eq5:4}) is obtained that
\begin{eqnarray*}
&a_0=-{\frac {\omega}3}+{\frac {m^2-2}3}\left({\frac {2\omega(3-\omega)}{m^4-m^2+1}}\right)^{\frac 12},a_1=0,\\
&{\quad}a_2=8c_4,c_2={\frac {m^2-2}8}\left({\frac {2\omega(3-\omega)}{m^4-m^2+1}}\right)^{\frac 12},
\end{eqnarray*}
which implies the Jacobian elliptic function solution of Eq.(\ref{eq5:1}) as
\begin{eqnarray}
\label{eq5:14}
&u(x,t)=-{\frac {\omega}3}-{\frac {m^2-2}3}\left({\frac {2\omega(3-\omega)}{m^4-m^2+1}}\right)^{\frac 12}
  -\left({\frac {2\omega(3-\omega)}{m^4-m^2+1}}\right)^{\frac 12}\dn^2(\eta,m),\nonumber\\
&\quad\eta={\frac 12}\left({\frac {\omega(3-\omega)}{2(m^4-m^2+1)}}\right)^{\frac 14}(x-{\omega}t),0<\omega<3.
\end{eqnarray}
\par\noindent
(4)\,For $c_2=0$,the Eq.(\ref{eq5:4}) solves
\begin{eqnarray*}
&a_0=-{\frac {\omega}3},a_1=0,a_2=8c_4,c_0={\frac {\omega(\omega-3)}{96c_4}},
\end{eqnarray*}
which gives the Jacobian elliptic function solution of Eq.(\ref{eq5:1}) as following
\begin{eqnarray}
\label{eq5:15}
&u(x,t)=-{\frac {\omega}3}+{\frac 23}\sqrt{6\omega(3-\omega)}
   \ds^2\left(\eta,{\frac {\sqrt{2}}2}\right),\nonumber\\
&{\quad}\eta={\frac 16}\left(54\omega(3-\omega)\right)^{\frac 14}(x-{\omega}t),0<\omega<3,\\
\label{eq5:16}
&u(x,t)=-{\frac {\omega}3}+{\frac {\varepsilon}3}\sqrt{6\omega(\omega-3)}\left[\ns(\eta,{\frac {\sqrt{2}}2})+\cs(\eta,{\frac {\sqrt{2}}2})\right]^2,\nonumber\\
&\quad\eta=384\left(6\omega(\omega-3)\right)^{\frac 14}(x-{\omega}t),
\omega<0\;\mbox{or}\;\omega>3.
\end{eqnarray}
\newline
{\bf D.} When $c_2=c_4=0$,the solution of Eq.(\ref{eq5:4}) is found to be
\begin{eqnarray*}
a_0=-{\frac {\omega}3},a_1=2c_3,a_2=0,c_1={\frac {\omega(3-\omega)}{6c_3}},
\end{eqnarray*}
which gives the Weierstrass elliptic function solution of Eq.(\ref{eq5:1}) as
\begin{eqnarray}
\label{eq5:17}
u(x,t)=2c_3\wp\left({\frac {\sqrt{c_3}}2}(x-{\omega}t),{\frac {2\omega(3-\omega)}{3c_3^2}},-{\frac {4c_0}{c_3}}\right).
\end{eqnarray}
\newline
{\bf E.} When $c_0=0$,we have
\newline
(1)\,From the condition of $F_{27}$ we obtain the solution of Eq.(\ref{eq5:4})
as following
\begin{eqnarray*}
&a_0=-{\frac {\omega}3}+{\frac {2m^2-1}3}\left({\frac {2\omega(3-\omega)}{m^4-m^2+1}}\right)^{\frac 12},a_1=4c_3,\\
&{\quad}a_2={\frac {2c_3^2}{m^2}}\left({\frac {2(m^4-m^2+1)}{\omega(3-\omega)}}\right)^{\frac 12},c_4={\frac {c_3^2}{4m^2}}\left({\frac {2(m^4-m^2+1)}{\omega(3-\omega)}}\right)^{\frac 12},
\end{eqnarray*}
from which we obtain the Jacobian snoidal wave solution of Eq.(\ref{eq5:1}) as
\begin{eqnarray}
\label{eq5:18}
&u(x,t)=-{\frac {\omega}3}-{\frac {m^2+1}3}
  \left({\frac {2\omega(3-\omega)}{m^4-m^2+1}}\right)^{\frac 12}
  +m^2\left({\frac {2\omega(3-\omega)}{m^4-m^2+1}}\right)^{\frac 12}\sn^2\left(\rho(x-{\omega}t,m\right),\nonumber\\
&{\quad}\rho=-{\frac 12}\left({\frac {\omega(3-\omega)}{2(m^4-m^2+1)}}\right)^{\frac 14},0<\omega<3.
\end{eqnarray}
(2)\,Under condition of $F_{28}$,the Eq.(\ref{eq5:4}) solves that
\begin{eqnarray*}
&a_0=-{\frac {\omega}3}+{\frac {m^2-1}3}\left({\frac {2\omega(3-\omega)}{m^4-m^2+1}}\right)^{\frac 12},a_1=4c_3,\\
&{\quad}a_2={\frac {2c_3^2}{m^2}}\left({\frac {2(m^4-m^2+1)}{\omega(3-\omega)}}\right)^{\frac 12},
c_4={\frac {c_3^2}{4m^2}}\left({\frac {2(m^4-m^2+1)}{\omega(3-\omega)}}\right)^{\frac 12},
\end{eqnarray*}
from which we obtain the Jacobian elliptic function solution of Eq.(\ref{eq5:1}) in the form
\begin{eqnarray}
\label{eq5:19}
&u(x,t)=-{\frac {\omega}3}-{\frac {m^2+1}3}
  \left({\frac {2\omega(3-\omega)}{m^4-m^2+1}}\right)^{\frac 12}+\left({\frac {2\omega(3-\omega)}{m^4-m^2+1}}\right)^{\frac 12}\ns^2\left(\rho(x-{\omega}t,m\right),\nonumber\\
&\quad\rho=-{\frac 12}\left({\frac {\omega(3-\omega)}{2(m^4-m^2+1)}}\right)^{\frac 14},0<\omega<3.
\end{eqnarray}
(3)\,The solution of Eq.(\ref{eq5:4}) induced by the condition of $F_{29}$ is that
\begin{eqnarray*}
&a_0=-{\frac {\omega}3}-{\frac {m^2-2}3}\left({\frac {2\omega(3-\omega)}{m^4-m^2+1}}\right)^{\frac 12},a_1=4c_3,\\
&{\quad}a_2=2c_3^2\left({\frac {2(m^4-m^2+1)}{\omega(3-\omega)}}\right)^{\frac 12},
c_4={\frac {c_3^2}{4}}\left({\frac {2(m^4-m^2+1)}{\omega(3-\omega)}}\right)^{\frac 12},
\end{eqnarray*}
which leads the Jacobian snoidal wave solution of Eq.(\ref{eq5:1}) as
\begin{eqnarray}
\label{eq5:20}
&u(x,t)=-{\frac {\omega}3}-{\frac {m^2+1}3}
  \left({\frac {2\omega(3-\omega)}{m^4-m^2+1}}\right)^{\frac 12}
  +m^2\left({\frac {2\omega(3-\omega)}{m^4-m^2+1}}\right)^{\frac 12}\sn^2\left(\rho(x-{\omega}t,m\right),\nonumber\\
&\quad\rho=-{\frac 12}\left({\frac {\omega(3-\omega)}{2(m^4-m^2+1)}}\right)^{\frac 14},0<\omega<3.
\end{eqnarray}
(4)\,Using the condition of $F_{30}$ we find the solution of Eq.(\ref{eq5:4}) as following
\begin{eqnarray*}
&a_0=-{\frac {\omega}3}-{\frac {m^2-2}3}\left({\frac {2\omega(3-\omega)}{m^4-m^2+1}}\right)^{\frac 12},a_1=4c_3,\\
&{\quad}a_2=2c_3^2\left({\frac {2(m^4-m^2+1)}{\omega(3-\omega)}}\right)^{\frac 12},c_4={\frac {c_3^2}{4}}\left({\frac {2(m^4-m^2+1)}{\omega(3-\omega)}}\right)^{\frac 12},
\end{eqnarray*}
from which we obtain the Jacobian elliptic function solution of Eq.(\ref{eq5:1}) in the form
\begin{eqnarray}
\label{eq5:21}
&u(x,t)=-{\frac {\omega}3}-{\frac {m^2+1}3}
  \left({\frac {2\omega(3-\omega)}{m^4-m^2+1}}\right)^{\frac 12}
  +\left({\frac {2\omega(3-\omega)}{m^4-m^2+1}}\right)^{\frac 12}\ns^2\left(\rho(x-{\omega}t,m\right),\nonumber\\
&\quad\rho=-{\frac 12}\left({\frac {\omega(3-\omega)}{2(m^4-m^2+1)}}\right)^{\frac 14},0<\omega<3.
\end{eqnarray}
(5)\,From the condition of $F_{31}$ we yields the solution of Eq.(\ref{eq5:4}) as following
\begin{eqnarray*}
&a_0=-{\frac {\omega}3}+{\frac {m^2+1}3}\left({\frac {2\omega(3-\omega)}{m^4-m^2+1}}\right)^{\frac 12},a_1=4c_3,\\
&{\quad}a_2={\frac {2c_3^2}{m^2}}\left({\frac {2(m^4-m^2+1)}{\omega(3-\omega)}}\right)^{\frac 12},c_4={\frac {c_3^2}{4m^2}}\left({\frac {2(m^4-m^2+1)}{\omega(3-\omega)}}\right)^{\frac 12},
\end{eqnarray*}
which gives the Jacobian snoidal wave solution of Eq.(\ref{eq5:1}) in the form
\begin{eqnarray}
\label{eq5:22}
&u(x,t)=-{\frac {\omega}3}-{\frac {m^2+1}3}
  \left({\frac {2\omega(3-\omega)}{m^4-m^2+1}}\right)^{\frac 12}
   +m^2\left({\frac {2\omega(3-\omega)}{m^4-m^2+1}}\right)^{\frac 12}
   \sn^2\left(\rho(x-{\omega}t,m\right),\nonumber\\
&\quad\rho={\frac 12}\left({\frac {\omega(3-\omega)}{2(m^4-m^2+1)}}\right)^{\frac 14},0<\omega<3.
\end{eqnarray}
(6)\,By the condition of $F_{32}$ we get the solution of Eq.(\ref{eq5:4}) as following
\begin{eqnarray*}
&a_0=-{\frac {\omega}3}-{\frac {m^2+1}3}\left({\frac {2\omega(3-\omega)}{m^4-m^2+1}}\right)^{\frac 12},a_1=4c_3,\\
&{\quad}a_2={\frac {2c_3^2}{m^2}}\left({\frac {2(m^4-m^2+1)}
   {\omega(3-\omega)}}\right)^{\frac 12},
   c_4={\frac {c_3^2}{4m^2}}\left({\frac {2(m^4-m^2+1)}{\omega(3-\omega)}}\right)^{\frac 12},
\end{eqnarray*}
which leads the Jacobian elliptic function solution of Eq.(\ref{eq5:1}) in the form
\begin{eqnarray}
\label{eq5:23}
&u(x,t)=\left[-{\frac {\omega}3}+{\frac {2m^2-1}3}
  \left({\frac {2\omega(3-\omega)}{m^4-m^2+1}}\right)^{\frac 12}\right]\nd^2\left(\rho(x-{\omega}t,m\right)\nonumber\\
&\quad+\left[{\frac {m^2\omega}{3}}+{\frac {m^2(m^2-2)}{3}}\left({\frac {2\omega(3-\omega)}{m^4-m^2+1}}\right)^{\frac 12}\right]\sd^2(\rho(x-{\omega}t),m),\nonumber\\
&\qquad\rho={\frac 12}\left({\frac {\omega(3-\omega)}{2(m^4-m^2+1)}}\right)^{\frac 14},0<\omega<3.
\end{eqnarray}
(7)\,The solution of Eq.(\ref{eq5:4}) induced by the condition of $F_{33}$ reads
\begin{eqnarray*}
&a_0=-{\frac {\omega}3}+{\frac {2m^2-1}3}\left({\frac {2\omega(3-\omega)}{m^4-m^2+1}}\right)^{\frac 12},a_1=4c_3,\\
&{\quad}a_2={\frac {2c_3^2}{m^2-1}}\left({\frac {m^4-m^2+1}{\omega(3-\omega)}}\right)^{\frac 12},c_4={\frac {c_3^2}{4(m^2-1)}}\left({\frac {m^4-m^2+1}{\omega(3-\omega)}}\right)^{\frac 12},
\end{eqnarray*}
from which we obtain the Jacobian elliptic function solution of Eq.(\ref{eq5:1}) as
\begin{eqnarray}
\label{eq5:24}
&u(x,t)=-{\frac {\omega}3}-{\frac {m^2-2}3}
  \left({\frac {2\omega(3-\omega)}{m^4-m^2+1}}\right)^{\frac 12}
  -\left({\frac {2\omega(3-\omega)}{m^4-m^2+1}}\right)^{\frac 12}
   \dn^2\left(\rho(x-{\omega}t,m\right),\nonumber\\
&{\quad}\rho=-{\frac 12}\left({\frac {\omega(3-\omega)}{2(m^4-m^2+1)}}\right)^{\frac 14},0<\omega<3.
\end{eqnarray}
(8)\,The condition of $F_{34}$ leads the solution of Eq.(\ref{eq5:4}) as following
\begin{eqnarray*}
&a_0=-{\frac {\omega}3}+{\frac {2m^2-1}3}\left({\frac {2\omega(3-\omega)}{m^4-m^2+1}}\right)^{\frac 12},a_1=4c_3,\\
&{\quad}a_2={\frac {2c_3^2}{m^2-1}}\left({\frac {2(m^4-m^2+1)}{\omega(3-\omega)}}\right)^{\frac 12},
c_4={\frac {c_3^2}{4(m^2-1)}}\left({\frac {2(m^4-m^2+1)}{\omega(3-\omega)}}\right)^{\frac 12},
\end{eqnarray*}
which gives the Jacobian elliptic function solution of Eq.(\ref{eq5:1}) in the form
\begin{eqnarray}
\label{eq5:25}
&u(x,t)=\left[-{\frac {\omega}3}+{\frac {2m^2-1}3}
  \left({\frac {2\omega(3-\omega)}{m^4-m^2+1}}\right)^{\frac 12}\right]nd^2\left(\rho(x-{\omega}t,m)\right)\nonumber\\
&\quad+\left[{\frac {m^2\omega}{3}}+{\frac {m^2(m^2-2)}{3}}\left({\frac {2\omega(3-\omega)}{m^4-m^2+1}}\right)^{\frac 12}\right]\sd^2(\rho(x-{\omega}t,m),\nonumber\\
&\qquad\rho=-{\frac 12}\left({\frac {\omega(3-\omega)}{2(m^4-m^2+1)}}\right)^{\frac 14},0<\omega<3.
\end{eqnarray}
(9)\,The solution of Eq.(\ref{eq5:4}) obtained by the condition of $F_{35}$ reads
\begin{eqnarray*}
&a_0=-{\frac {\omega}3}-{\frac {m^2-2}3}\left({\frac {2\omega(3-\omega)}{m^4-m^2+1}}\right)^{\frac 12},a_1=4c_3,\\
&{\quad}a_2={\frac {2c_3^2}{m^2-1}}\left({\frac {2(m^4-m^2+1)}{\omega(3-\omega)}}\right)^{\frac 12},c_4={\frac {c_3^2}{4(m^2-1)}}\left({\frac {2(m^4-m^2+1)}{\omega(3-\omega)}}\right)^{\frac 12},
\end{eqnarray*}
from which we obtain the Jacobian elliptic function solution of Eq.(\ref{eq5:1}) as
\begin{eqnarray}
\label{eq5:26}
&u(x,t)=\left[-{\frac {\omega}3}-{\frac {m^2-2}3}
  \left({\frac {2\omega(3-\omega)}{m^4-m^2+1}}\right)^{\frac 12}\right]\nc^2\left(\rho(x-{\omega}t,m)\right)\nonumber\\
&\quad+\left[{\frac {\omega}{3}}-{\frac {2m^2-1}{3}}\left({\frac {2\omega(3-\omega)}{m^4-m^2+1}}\right)^{\frac 12}\right]\sc^2(\rho(x-{\omega}t,m),\nonumber\\
&\quad\rho=-{\frac 12}\left({\frac {\omega(3-\omega)}{2(m^4-m^2+1)}}\right)^{\frac 14},0<\omega<3.
\end{eqnarray}
(10)\,By the condition of $F_{36}$ we get the solution of Eq.(\ref{eq5:4}) as following
\begin{eqnarray*}
&a_0=-{\frac {\omega}3}-{\frac {m^2-2}3}\left({\frac {2\omega(3-\omega)}{m^4-m^2+1}}\right)^{\frac 12},a_1=4c_3,\\
&{\quad}a_2={\frac {2c_3^2}{1-m^2}}\left({\frac {2(m^4-m^2+1)}
   {\omega(3-\omega)}}\right)^{\frac 12},c_4={\frac {c_3^2}{4(1-m^2)}}\left({\frac {2(m^4-m^2+1)}{\omega(3-\omega)}}\right)^{\frac 12},
\end{eqnarray*}
which leads the Jacobian elliptic function solution of Eq.(\ref{eq5:1}) in the form
\begin{eqnarray}
\label{eq5:27}
&u(x,t)=-{\frac {\omega}3}-{\frac {m^2+1}3}
  \left({\frac {2\omega(3-\omega)}{m^4-m^2+1}}\right)^{\frac 12}+\left({\frac {2\omega(3-\omega)}{m^4-m^2+1}}\right)^{\frac 12}\ns^2(\rho(x-{\omega}t),m),\nonumber\\
&\quad\rho=-{\frac 12}\left({\frac {\omega(3-\omega)}{2(m^4-m^2+1)}}\right)^{\frac 14},0<\omega<3.
\end{eqnarray}
(11)\,The condition of $F_{37}$ gives the solution of Eq.(\ref{eq5:4}) as following
\begin{eqnarray*}
&a_0=-{\frac {\omega}3}-{\frac {m^2+1}3}\left({\frac {2\omega(3-\omega)}{m^4-m^2+1}}\right)^{\frac 12},a_1=4c_3,\\
&{\quad}a_2=2c_3^2\left({\frac {2(m^4-m^2+1)}{\omega(3-\omega)}}\right)^{\frac 12},
c_4={\frac {c_3^2}{4}}\left({\frac {2(m^4-m^2+1)}{\omega(3-\omega)}}\right)^{\frac 12},
\end{eqnarray*}
which gives the Jacobian snoidal wave solution of Eq.(\ref{eq5:1}) in the form
\begin{eqnarray}
\label{eq5:28}
&u(x,t)=-{\frac {\omega}3}-{\frac {m^2+1}3}
  \left({\frac {2\omega(3-\omega)}{m^4-m^2+1}}\right)^{\frac 12}
  +m^2\left({\frac {2\omega(3-\omega)}{m^4-m^2+1}}\right)^{\frac 12}\sn^2(\rho(x-{\omega}t),m),\nonumber\\
&{\quad}\rho={\frac 12}\left({\frac {\omega(3-\omega)}{2(m^4-m^2+1)}}\right)^{\frac 14},0<\omega<3.
\end{eqnarray}
(12)\,The solution of Eq.(\ref{eq5:4}) obtained by using the condition of $F_{38}$ reads
\begin{eqnarray*}
&a_0=-{\frac {\omega}3}-{\frac {m^2+1}3}\left({\frac {2\omega(3-\omega)}{m^4-m^2+1}}\right)^{\frac 12},a_1=4c_3,\\
&{\quad}a_2=2c_3^2\left({\frac {2(m^4-m^2+1)}{\omega(3-\omega)}}\right)^{\frac 12},
c_4={\frac {c_3^2}{4}}\left({\frac {2(m^4-m^2+1)}{\omega(3-\omega)}}\right)^{\frac 12},
\end{eqnarray*}
from which we obtain the Jacobian elliptic function solution of Eq.(\ref{eq5:1}) as
\begin{eqnarray}
\label{eq5:29}
&u(x,t)=\left[-{\frac {\omega}3}+{\frac {2m^2-1}3}
  \left({\frac {2\omega(3-\omega)}{m^4-m^2+1}}\right)^{\frac 12}\right]\nd^2\left(\rho(x-{\omega}t,m)\right)\nonumber\\
&\quad+\left[{\frac {m^2\omega}{3}}+{\frac {m^2(m^2-2)}{3}}\left({\frac {2\omega(3-\omega)}{m^4-m^2+1}}\right)^{\frac 12}\right]\sd^2(\rho(x-{\omega}t,m),\nonumber\\
&\quad\rho={\frac 12}\left({\frac {\omega(3-\omega)}{2(m^4-m^2+1)}}\right)^{\frac 14},0<\omega<3.
\end{eqnarray}
\section{Concluding remarks}
\label{sec:6}
We constructed the BT and the SF of two kinds of sub-equations for Eq.(\ref{eq1:1})
in terms of the indirect mapping method for simplicity.As a matter of fact,
we can establish the BT and the SF for most of those sub-equations for
Eq.(\ref{eq1:1}) by the elementary integration method which were
not considered here and will be reported elsewhere.It is also clear that
the new solutions of Eq.(\ref{eq1:1}) obtained from the BT or SF can also be
used to find new solutions of NLEEs which is under consideration.But an
open problem as how to construct multiple traveling wave solutions of NLEEs
from the solutions of NLEEs obtained by using the auxiliary equations
is to be solved.\par
It should be noticed that the four lemmas presented in Sec.\ref{sec:3}
can be widely used to prove the equivalence relations of numerous solutions
for other auxiliary equations.At the same time,the four lemmas can also
be applied to prove the equivalence relations of some direct methods.
Therefore,the proof of the equivalence relations of solutions for the auxiliary
equations is more important than that of discussing the equivalence relations
of those obtained solutions for a given nonlinear equation.This is the reason
why we studied the equivalence relations of solutions for Eq.(\ref{eq1:1})
here in detail.\par
In Sec.\ref{sec:5} we obtained some exact traveling wave solutions of wave
speed defined on the region $\omega\in(0,3)$ and
$\omega\in(-\infty,0)\cup(3,+\infty)$ for mCH equation that was not considered
and obtained before so they are the new solutions for mCH equation.Although
some of them are same in form but they are obtained by using different conditions
and therefore they can be regarded as new solutions as well.In the limit of
$m\rightarrow{1}$,the solitary wave solutions for mCH equation can be obtained
from the Jacobian elliptic function solutions constructed in Sec.\ref{sec:5},
and which were omitted.


\begin{thebibliography}{99}
\bibitem{RefJ-1}
S.K.Liu,Z.T. Fu,S.D.Liu,and Q. Zhao,``Jacobi elliptic function expansion method
   and periodic wave solutions of nonlinear wave equations," Phys. Lett. A
  \textbf{289},69--74 (2001).
\bibitem{RefJ-2}
M.L.Wang and X.Z.Li,``Applications of F--expansion to periodic wave solutions
   for a new Hamiltonian amplitude equation," Chaos, Solitons and Fractals
   \textbf{24},1257--1268 (2005).
\bibitem{RefJ-3}
Sirendaoreji and J.Sun,``Auxiliary equation method for solving nonlinear partial
   differential equations," Phys. Lett. A \textbf{309},387--396 (2003).
\bibitem{RefJ-4}
E.G.Fan,``Uniformly constructing a series of explicit exact solutions to nonlinear
  equations in mathematical physics," Chaos,Solitons and Fractals
  \textbf{16},819--839 (2003).
\bibitem{RefJ-5}
E.Yomba,``The extended Fan's sub-equation method and its application to KdV-MKdV,
  BKK and variant Boussinesq equations,"  Phys. Lett. A \textbf{336},463--476 (2005).
\bibitem{RefJ-6}
A.A.Soliman and M.A.Abdou,``Exact travelling wave solutions of nonlinear partial
   differential equations," Chaos, Solitons and Fractals \textbf{32},808--815 (2007).
\bibitem{RefB-1}
S.K.Liu and S.D.Liu,``Nonlinear Equations in Physics," 32--53.
   Peking University Press, Peking (2000).
\bibitem{RefB-2}
E.G.Fan,``Integrable Systems and Computer Algebra," 29--55.
   Science Press, Peking (2004).
\bibitem{RefB-3}
Z.B.Li,``Traveling Wave Solutions of Nonlinear Equations in Mathematical Physics,"
   147--153.Science Press, Peking (2007).
\bibitem{RefB-4}
Y.C.Guo,``Nonlinear Partial Differential Equations," 223--226.
   Tsinghua University Press, Peking (2008).
\bibitem{RefJ-7}
N.Taghizadeh,Q.Zhou,M.Ekici,and M.Mirzazadeh,``Soliton solutions for Davydov solitons in
    $\alpha$--helix proteins," Superlattices and Microstructures \textbf{102},323--341 (2017).
\bibitem{RefJ-8}
 M.M.El--Borai,H.M.El--Owaidy,H.M.Ahmed,and A.H.Arnous,``Exact and soliton
    solutions to nonlinear transmission line model," Nonlinear Dyn.
    \textbf{87}, 767--773 (2017).
\bibitem{RefJ-9}
Sirendaoreji,``New exact traveling wave solutions for the Kawahara and the modified
              Kawahara equations," Chaos, Solitons and Fractals \textbf{19},
              147--150 (2004).
\bibitem{RefJ-10}
Sirendaoreji,``Auxiliary equation method and new solutions of Klein-Gordon equations,"
    Chaos, Solitons and Fractals \textbf{31}: 943--950 (2007).
\bibitem{RefJ-11}
E.Yomba,``The sub-ODE method for finding exact travelling wave solutions of generalized
    nonlinear Camassa-Holm,and generalized nonlinear Schr\"odinger equations,"
    Phys. Lett. A \textbf{372},215--222 (2008).
\bibitem{RefJ-12}
M.A.Abdou,``A generalized auxiliary equation method and its applications,"
    Nonlinear Dyn. \textbf{52}: 95--102 (2008).
\bibitem{RefJ-13}
N.Cheemaa and M.Younis,``New and more exact traveling wave solutions to integrable
    (2+1)-dimensional Maccari system," Nonlinear Dyn. \textbf{83}:1395--1401 (2016).
\bibitem{RefJ-14}
Sirendaoreji,``Exact travelling wave solutions for four forms of nonlinear
     Klein--Gordon equations," Phys. Lett. A \textbf{363}: 440--447 (2007).
\bibitem{RefJ-15}
H.Triki,H.Leblond,and D.Mihalache,``Soliton solutions of nonlinear
     diffusion--reaction--type equations with time--dependent
     coefficients accounting for long--range diffusion," Nonlinear Dyn.
     \textbf{86}:2115--2126 (2016).
\bibitem{RefJ-16}
J.P.Yu,D.S.Wang,,Y.L.Sun,and S.P.Wu,``Modified method of simplest equation
   for obtaining exact solutions of the Zakharov--Kuznetsov equation, the modified
   Zakharov-Kuznetsov equation, and their generalized forms," Nonlinear Dyn.
   \textbf{85},2449--2465 (2016).
\bibitem{RefJ-17}
X.Z.Li and M.L.Wang,``A sub--ODE method for finding exact solutions of a generalized
   KdV--mKdV equation with high-order nonlinear terms," Phys.Lett. A \textbf{361},
   115--118 (2007).
\bibitem{RefJ-18}
C.S.Liu,``Applications of complete discrimination system for polynomial for
   classifications of traveling wave solutions to nonlinear differential equations,"
   Comput.Phys. Commun. \textbf{181},317--324 (2010).
\bibitem{RefJ-19}
J.Y.Hu,``Classification of single traveling wave solutions to the nonlinear
   dispersion Drinfel'd--Sokolov system," Appl. Math. Comput.
   \textbf{219}, 2017--2025 (2012).
\bibitem{RefJ-20}
Taogetusang,``Several auxiliary equations and infinite sequence exact solutions
   to nonlinear evolution equations," Acta Phys. Sin. \textbf{60},050201:1--13 (2011).
\bibitem{RefJ-21}
X.Zeng and X.L.Yong,``A new mapping method and its applications to nonlinear
   partial differential equations," Phys. Lett. A \textbf{372},6602--6607 (2008).
\bibitem{RefJ-22}
S.Zhang,``A generalized auxiliary equation method and its application to
   (2 + 1)--dimensional Korteweg--de Vries equations," Comput. Math. Appl.
   \textbf{54},1028--1038 (2007).
\bibitem{RefJ-23}
S.Zhang and T.C.Xia,``A generalized auxiliary equation method and its
    application to (2+1)-dimensional asymmetric Nizhnik--Novikov--Vesselov equations,"
    J. Phys. A: Math. Theor. \textbf{40}: 227--248 (2007).
\bibitem{RefJ-24}
D.J.Hunag,D.S.Li and H.Q.Zhang,``Explicit and exact travelling
    wave solutions for the generalized derivative Shr\"{o}dinger equation,"
    Chaos, Solitons and Fractals \textbf{31},586--593 (2007).
\bibitem{RefJ-25}
G.Q.Xu,``Extended auxiliary equation method and its applications to
    three generalized NLS equations," Abstract and Appl. Anal.
    \textbf{2014}, 1--7 (2014).
\bibitem{RefJ-26}
E.M.E.Zayed and K.A.E.Alurrfi,``Extended auxiliary equation method and its
   applications for finding the exact solutions for a class of nonlinear
   Schr\"oinger--type equations," Appl. Math. Comput. \textbf{289},111-131 (2016).
\bibitem{RefJ-27}
Z.Pinar and T.\"Qzis,``An observation on the periodic solutions to nonlinear
   physical models by means of the auxiliary equation with a sixth-degree
   nonlinear term. Commun. Nonl. Sci. Numer. Simulat. \textbf{18}, 2177--2187 (2013).
\bibitem{RefJ-28}
Sirendaoreji,``Unified Riccati equation expansion method and its application to
   two new classes of Benjamin--Bona--Mahony equations. Nonlinear Dyn.
   \textbf{89},333--344 (2017)
\bibitem{RefJ-29}
C.P.Liu and X.P.Liu,``A note on the auxiliary equation method for solving nonlinear
   partial differential equations," Phys. Lett. A \textbf{348},222--227 (2006).
\bibitem{RefJ-30}
X.P.Liu and C.P.Liu,``The relationship among the solutions of two auxiliary ordinary
   differential equations," Chaos Solitons and Fractals \textbf{39},1915--1919 (2009).
\bibitem{RefJ-31}
C.D.Tian,``The rational solutions for a kind of mCH equation," J. Qilu Norm. Univ.
   \textbf{26},119--123 (2011).
\bibitem{RefJ-32}
X.L.Yang and J.S.Tan,``New travelling wave solutions for combined KdV--mKdV equation and
   (2+1)--dimensional Broer--Kaup--Kupershmidt system," Chin.Phys. \textbf{16},310--317 (2007).
\bibitem{RefJ-33}
A.H.Lu and X.K.Shao,``Using the elliptical auxiliary equation method to obtain
   exact solutions of the variable coefficients Boussinesq equation," J. Chongqing
   Univ. Art. Sci.\textbf{33},11--13 (2014).
\bibitem{RefJ-34}
Taogetusang and Sirendaoreji,`` New type of exact solitary wave solutions for
   dispersive long--wave equation and Benjamin equation," Acta Physica Sinica \textbf{55},3246--3254 (2006).
\bibitem{RefJ-35}
A.M.Wazwaz,``Solitary wave solutions for modified forms of Degasperis--Procesi and
   Camassa--Holm equations," Phys. Lett. A, \textbf{352},500-504 (2006).
\bibitem{RefJ-36}
A.M.Wazwaz,``New solitary wave solutions to the modified forms of Degasperis-Procesi
    and Camassa-Holm equations," Appl. Math. Comput. \textbf{186},130--141 (2007).
\bibitem{RefJ-37}
Q.D.Wang and M.Y.Tang,``New exact solutions for two nonlinear equations," Phys.
    Lett. A,\textbf{372},2995--3000 (2008).
\bibitem{RefJ-38}
S.X.Liang and D.J.Jeffrey,``New traveling wave solutions to modified CH and DP
    equations," Comp.Phys.Commun.\textbf{180},1429--1433 (2009).
\end{thebibliography}
\end{document}